\documentclass{emulateapj}
\usepackage{natbib}
\usepackage{lscape}
\usepackage{pdflscape}
\usepackage{amssymb}
\usepackage{graphicx}
\usepackage{rotating}
\def\CIVdblt{{\rm C~}\kern 0.1em{\sc iv}~$\lambda\lambda 1548, 1550$}
\def\MgIIdblt{{\rm Mg~}\kern 0.1em{\sc ii}~$\lambda\lambda 2796, 2803$}
\def\NVdblt{{\rm N}\kern 0.1em{\sc v}~$\lambda\lambda 1238, 1242$}  
\def\OVIdblt{{\rm O}\kern 0.1em{\sc vi}~$\lambda\lambda 1032, 1038$}
\def\SiIVdblt{{\rm Si~}\kern 0.1em{\sc iv}~$\lambda\lambda1394, 1403$}
\def\AlIIIdblt{{\rm Al~}\kern 0.1em{\sc iii}~$\lambda\lambda1855,1863$}
\def\FeIIdblt{{\rm Fe~}\kern 0.1em{\sc ii}~$\lambda\lambda 2383, 2600$}
\def\NeVIIIdblt{{\rm Ne~}\kern 0.1em{\sc viii}~$\lambda\lambda 770, 780$}

\def\NeVIII{\hbox{{\rm Ne~}\kern 0.1em{\sc viii}}}
\def\OI{\hbox{{\rm O~}\kern 0.1em{\sc i}}}
\def\OII{\hbox{{\rm O~}\kern 0.1em{\sc ii}}}
\def\OIII{\hbox{{\rm O~}\kern 0.1em{\sc iii}}}
\def\OIV{\hbox{{\rm O~}\kern 0.1em{\sc iv}}}
\def\OVI{\hbox{{\rm O~}\kern 0.1em{\sc vi}}}
\def\OVII{\hbox{{\rm O~}\kern 0.1em{\sc vii}}}
\def\OVIII{\hbox{{\rm O~}\kern 0.1em{\sc viii}}}
\def\NIII{\hbox{{\rm N~}\kern 0.1em{\sc iii}}}
\def\NIV{\hbox{{\rm N~}\kern 0.1em{\sc iv}}}
\def\NVII{\hbox{{\rm N~}\kern 0.1em{\sc vii}}}
\def\CIII{\hbox{{\rm C~}\kern 0.1em{\sc iii}}}
\def\SiIII{\hbox{{\rm Si~}\kern 0.1em{\sc iii}}}
\def\SVI{\hbox{{\rm S~}\kern 0.1em{\sc vi}}}
\def\NeIX{\hbox{{\rm Ne~}\kern 0.1em{\sc ix}}}

\def\AlII{\hbox{{\rm Al~}\kern 0.1em{\sc ii}}}
\def\AlIII{\hbox{{\rm Al~}\kern 0.1em{\sc iii}}}
\def\CaI{\hbox{{\rm Ca}\kern 0.1em{\sc i}}}
\def\CaII{\hbox{{\rm Ca}\kern 0.1em{\sc ii}}}
\def\CrII{\hbox{{\rm Cr}\kern 0.1em{\sc ii}}}
\def\CII{\hbox{{\rm C~}\kern 0.1em{\sc ii}}}
\def\CIII{\hbox{{\rm C~}\kern 0.1em{\sc iii}}}
\def\CIV{\hbox{{\rm C~}\kern 0.1em{\sc iv}}}
\def\CV{\hbox{{\rm C}\kern 0.1em{\sc v}}}
\def\H{\hbox{{\rm H}}}
\def\HI{\hbox{{\rm H~}\kern 0.1em{\sc i}}}
\def\HII{\hbox{{\rm H~}\kern 0.1em{\sc ii}}}
\def\Lya{\hbox{{\rm Ly}\kern 0.1em$\alpha$}}
\def\Lyb{\hbox{{\rm Ly}\kern 0.1em$\beta$}}
\def\Lyg{\hbox{{\rm Ly}\kern 0.1em$\gamma$}}
\def\Lyth{\hbox{{\rm Ly}\kern 0.1em$\theta$}}
\def\Lyfive{\hbox{{\rm Ly}\kern 0.1em$5$}}
\def\Lysix{\hbox{{\rm Ly}\kern 0.1em$6$}}
\def\Lyseven{\hbox{{\rm Ly}\kern 0.1em$7$}}
\def\Lyeight{\hbox{{\rm Ly}\kern 0.1em$8$}}
\def\Lynine{\hbox{{\rm Ly}\kern 0.1em$9$}}
\def\Lyten{\hbox{{\rm Ly}\kern 0.1em$10$}}
\def\HeI{\hbox{{\rm He}\kern 0.1em{\sc i}}}
\def\HeII{\hbox{{\rm He}\kern 0.1em{\sc ii}}}
\def\FeI{\hbox{{\rm Fe~}\kern 0.1em{\sc i}}}
\def\FeII{\hbox{{\rm Fe~}\kern 0.1em{\sc ii}}}
\def\FeIII{\hbox{{\rm Fe~}\kern 0.1em{\sc iii}}}
\def\MnII{\hbox{{\rm Mn}\kern 0.1em{\sc ii}}}
\def\MgI{\hbox{{\rm Mg~}\kern 0.1em{\sc i}}}
\def\MgII{\hbox{{\rm Mg~}\kern 0.1em{\sc ii}}}
\def\MgIII{\hbox{{\rm Mg~}\kern 0.1em{\sc iii}}}
\def\MgIV{\hbox{{\rm Mg~}\kern 0.1em{\sc iv}}}
\def\NaI{\hbox{{\rm Na}\kern 0.1em{\sc i}}}
\def\NV{\hbox{{\rm N~}\kern 0.1em{\sc v}}}
\def\NII{\hbox{{\rm N~}\kern 0.1em{\sc ii}}}
\def\NIII{\hbox{{\rm N~}\kern 0.1em{\sc iii}}}
\def\OVI{\hbox{{\rm O~}\kern 0.1em{\sc vi}}}
\def\OV{\hbox{{\rm O~}\kern 0.1em{\sc v}}}
\def\OVII{\hbox{{\rm O~}\kern 0.1em{\sc vii}}}
\def\SiII{\hbox{{\rm Si~}\kern 0.1em{\sc ii}}}
\def\SiIII{\hbox{{\rm Si~}\kern 0.1em{\sc iii}}}
\def\SiIV{\hbox{{\rm Si~}\kern 0.1em{\sc iv}}}
\def\SII{\hbox{{\rm S}\kern 0.1em{\sc ii}}}
\def\SIII{\hbox{{\rm S}\kern 0.1em{\sc iii}}}
\def\SIV{\hbox{{\rm S}\kern 0.1em{\sc iv}}}
\def\TiII{\hbox{{\rm Ti}\kern 0.1em{\sc ii}}}
\def\ZnII{\hbox{{\rm Zn}\kern 0.1em{\sc ii}}}
\newcommand{\kms}{\hbox{km~s$^{-1}$}}
\newcommand{\cmsq}{\hbox{cm$^{-2}$}}
\newcommand{\cc}{\hbox{cm$^{-3}$}}
\def\kms{\hbox{km~s$^{-1}$}}      
\def\cmsq{\hbox{cm$^{-2}$}}
\def\cc{\hbox{cm$^{-3}$}}
\def\etal{et~al.\ }

\begin{document}

\title{Highly Ionized Plasma in the Halo of a \\ Luminous Spiral Galaxy near $z = 0.225$\altaffilmark{1}}
\author{Anand Narayanan\altaffilmark{2}, Blair D. Savage\altaffilmark{2}, Bart P. Wakker\altaffilmark{2}}

\altaffiltext{1}{Based on observations with the NASA/ESA {\it Hubble Space Telescope}, obtained at the Space Telescope Science Institute, which is operated by the Association of Universities for Research in Astronomy, Inc., under NASA contract NAS 05-26555, and the NASA-CNES/ESA {\it Far Ultraviolet Spectroscopic
Explorer} mission, operated by the Johns Hopkins University, supported by NASA contract NAS 05-32985.}
\altaffiltext{2}{Department of Astronomy, University of Wisconsin-Madison, Email: anand, savage, wakker@astro.wisc.edu}
\subjectheadings{galaxies: halos, intergalactic medium, quasars: absorption lines, quasars: individual: H 1821+643, ultraviolet: general}
\submitted{Accepted for publication in The Astrophyiscal Journal}

\begin{abstract}
We present analyses of the physical conditions in the $z(\OVI) = 0.22496$ and $z(\OVI) = 0.22638$ multi-phase absorption systems detected in the ultraviolet HST/{\it STIS} and {\it FUSE} spectra of the quasar H~$1821+643$ ($m_V = 14.2, z_{em} = 0.297$). Both absorbers are likely associated with the extended halo of a $\sim 2L_B^*$ Sbc-Sc galaxy situated at a projected distance of  $\sim 116~h_{71}^{-1}$~kpc from the sight line. The $z = 0.22496$ absorber is detected in {\CII}, {\CIII}, {\CIV}, {\OIII}, {\OVI}, {\SiII}, {\SiIII} and {\HI} ({\Lya} - Ly$\theta$) at $> 3\sigma$ significance. The components of {\SiIII} and {\SiII} are narrow with implied temperatures of $T \lesssim 3 \times 10^4$~K. The low and intermediate ions in this absorber are consistent with an origin in a $T \sim 10^4$~K photoionized gas with [Si/H] and [C/H] of $\sim -0.6$~dex. In contrast, the broader {\OVI} absorption is likely produced in collisionally ionized plasma under nonequilibrium conditions. The $z(\OVI) = 0.22638$ system has broad {\Lya} (BLA) and {\CIII} absorption offset by $v = -53$~{\kms} from {\OVI}. The {\HI} and {\CIII} line widths for the BLA imply $T = 1.1 \times 10^5$~K. For non-equilibrium cooling we obtain [C/H] $\sim -1.5$~dex and $N(\H) = 3.2 \times 10^{18}$~{\cmsq} in the BLA. The {\OVI}, offset from the BLA with no detected {\HI} or {\CIII}, is likely collisionally ionized at $T \sim 3 \times 10^5$~K. From the observed multiphase properties and the proximity to a luminous galaxy, we propose that the $z=0.22496$ absorber is an extragalactic analog of a highly ionized Galactic HVC, in which the {\OVI} is produced in transition temperature plasma ($T \sim 10^5$~K) at the interface layers between the  {\it warm} ($T < 5 \times 10^4$~K) HVC gas phase and the {\it hot} ($T \gtrsim 10^6$~K) coronal halo of the galaxy. The $z=0.22638$ {\OVI} - BLA absorber could be tracing a cooling condensing fragment in the nearby galaxy's {\it hot} gaseous halo. 

\end{abstract}

\section{Introduction}

Quasar absorption line studies as well as galaxy formation models have shown that the baryons in the extended halos of galaxies exist in the form of structures with different masses, spatial scales, densities and temperatures. In the Milky Way, the multiphase composition of the halo is evident from the distribution of numerous neutral and {\it warm} \footnote {Throughout this paper we use the terms {\it warm} and {\it hot} in a manner that is consistent with the traditional definition of these terms in ISM astronomy. Thus, {\it warm} refers to gas with $T \sim (0.3 - 3) \times 10^4$~K, and {\it hot} refers to gas with $T > 10^6$~K. For intermediate temperatures of $T  \sim (0.5 - 10) \times 10^5$~K, we use the term {\it transition temperature} to covey that gas in this unstable temperature regime is likely not in a state of equilibrium.} high velocity gas clouds (HVCs) pressure supported by the {\it hot} and diffuse corona \citep{sembach03, collins04, fox04}. The sources of high velocity gas surrounding the Milky Way include tidally stripped gas during interactions with satellite galaxies, accreting gas from the IGM, galactic scale outflows, and fragmentations from the cooling of a {\it hot} halo \citep[see reviews by ][]{wakker97,richter06a}. Detecting multiphase gas in the gaseous halos of external galaxies provides an opportunity to trace these varied processes at higher redshifts. The multiphase nature of the absorber can only be fully understood through a combination of line diagnostics from low, intermediate and high ions. Close resemblance of the physical properties with the high velocity gas in the Galactic halo can be a crucial pointer towards the specific nature of these higher redshift systems. 

Among the high ions, {\OVIdblt}~{\AA}\footnote{We use the oscillator strengths and the rest-wavelengths of electronic transitions from the atomic data given in Verner {\etal} (1994, 1996) for $\lambda < 912$~{\AA} and \citet{morton00, morton03}. The wavelengths are given as vacuum wavelengths rounded to the nearest natural number.} lines have been of particular importance for tracing interstellar gas with $T \sim (1 -3) \times 10^5$~K, due to the high relative abundance of oxygen and the large oscillator strength of the doublet lines ($f_{1032} = 0.133, f_{1038} = 0.067$). Ultraviolet spectroscopic observations of sight lines towards extragalactic objects have detected {\OVIdblt}~{\AA} absorption associated with many of the HVCs in the halo of the Milky Way. The {\it FUSE} survey of \citet{sembach03} found $\sim 60 - 85$\% of the sky covered by these high velocity {\OVI} absorbing clouds. Some of this {\OVI} is drawn from the same population as the neutral HVCs with $N(\HI) \gtrsim 2 \times 10^{18}$~{\cmsq} and thus is also detected in 21-cm radio emission above this column density limit. This neutral population has a sky covering fraction of $\sim 30$\%. The more highly ionized HVCs are detected through their absorption in the optical and UV spectra of background sources. Their non-detection in 21-cm emission constrains the {\HI} column density to $N(\HI) < 2 \times 10^{18}$~{\cmsq} \citep{sembach99, wakker03, collins04, ganguly05}.  The {\OVI} is an excellent tracer of collisionally ionized gas since $E > 114$~eV energies are required for its production from lower ionization stages. The observational constraints set by the ionic column density ratio of {\OVI} with low/intermediate ions such as {\SiII}, {\CII}, {\SiIII}, {\CIII} and other high ions such as {\CIV}, {\SiIV} and {\NV} indicate that the ionization process in the high velocity gas phase traced by {\OVI} is dominated by collisions of electrons with ions in a plasma with $T \sim 2 \times 10^5$~K. The gas at this temperature is susceptible to strong radiative cooling, with the {\OVI} ion also acting as a major cooling agent \citep{sutherland93}. In Galactic HVCs, the transition temperature phase with $T \sim (1 - 3)~\times~10^5$~K forms from the interaction of {\it warm} photoionized gas with the {\it hot} coronal halo, and is described by a temperature intermediate to the {\it warm} and {\it hot} phases \citep{sembach03, fox04, fox05}. 

Our current knowledge on the physical state of the highly ionized HVCs in the Galactic halo provides useful indicators for understanding the nature of {\OVI} absorption line systems detected at higher redshift. The redshift number density $dN/dz \sim 15$ for absorbers with $W_r(\OVI~\lambda 1032) > 30$~m{\AA} at $z < 0.5$ determined from observations \citep{tripp08, danforth08} and supported by cosmological hydrodynamic simulations \citep{tumlinson05, cen06, oppenheimer09} indicates a high frequency of incidence for {\OVI} absorbers at low-$z$. Similarities in physical conditions with Galactic HVCs would suggest that at least some fraction of the population of the {\OVI} absorbers are tracing high velocity gas in external galaxies.  Evidence is emerging from absorber-galaxy pair studies in quasar fields as well as more general correlation studies between absorbers and databases of galaxies from wide field surveys in favor of {\OVI} absorption preferentially tracing gas in the immediate vicinity (impact parameter, $\rho \lesssim 500~h_{71}^{-1}$~kpc) of one or more galaxies \citep{sembach04, tumlinson05, stocke06, tripp06, wakker09, chen09}. Whether the absorption in all those cases is arising from a distinct gaseous structure embedded within the galaxy's halo, from a more diffuse gaseous envelope surrounding the galaxy or from an intergalactic filament networking the individual galaxies is not always evident. In a subset of nearby {\OVI} absorbers detected at high velocities with respect to the LSR, the ionization properties, galactocentric distances and location in the region of the sky surrounding the Milky Way are all consistent with an origin in the general Local Group environment rather than the Galactic halo \citep{sembach03}. It is therefore possible that some of the {\OVI} absorbers detected at higher redshifts could also be tracing clouds within  a group environment rather than the gaseous halo of one of the galaxies in the group. A detailed understanding of the physical conditions and metallicity in each absorber might help to distinguish its origin and location.

In this paper we describe the astrophysical nature of  two multiphase absorption systems detected along the sight line to the bright quasar H~$1821+643$, that were previously reported in \citet{savage98}, \citet{tripp00} and \citet{tripp08}. The two absorption systems are within $\Delta v \sim 350$~{\kms} of each other. Ground based imaging observations of this quasar field have identified galaxies in the vicinity of these absorbers \citep{schneider92, savage98, tripp98a}. The organization of this paper is as follows. In Sec. 2 we provide details on the observations. Secs. 3 and 4 describe the observed properties of the two absorption systems. Ground based imaging observations of the H~$1821+643$ field and information on the galaxy identified in the vicinity of the absorbers are given in Sec. 5. Detailed investigations of the various ionization scenarios in each absorption system is considered in Sec. 6, followed by explanations on the astrophysical nature of these absorbers (Sec. 7). The main results are summarized in the last section. 

 \section{Observations}

In our analysis, we use {\it STIS} and {\it FUSE} ultraviolet spectra of the quasar H~$1821+643$. The {\it STIS} observations include the higher resolution (FWHM $\sim 7$~{\kms}) FUV echelle E140M with the 0.2\arcsec~$\times$~0.06\arcsec~slit, and the lower resolution G230M (FWHM $\sim 30$~{\kms}) grating with the 52\arcsec~$\times$~0.05\arcsec~slit spectra that cover the wavelength intervals 1150 - 1730~{\AA} and 1840 - 1930~{\AA} respectively (Prop. ID: 8165). The details of the observations are given in \citet{tripp00} and \citet{tripp08}. The {\it STIS} spectra combines a total integration of 50.932 ks in the E140M and 25.236 ks in the G230M gratings. The co-added E140M spectrum has a mean $S/N \sim 18$ per 7~{\kms} resolution element, and the G230M has a mean $S/N \sim 26$ per 30~{\kms} resolution element. 

The {\it FUSE} spectrum spans the $912 - 1187$~{\AA} wavelength range and thus covers the higher order Lyman series lines as well as a few metal lines for the absorption systems discussed in this paper. The spectral resolution is $\sim 20$~{\kms}. A velocity shift was applied to the exposures in each detector segment so that the spectral features are correctly aligned \citep[see ][for details]{wakker03, wakker06}. The magnitude of this shift was determined by fitting the velocity centroids of low-ion absorption components in the FUSE spectrum and aligning them with the respective low-ion components in the STIS spectrum. The final spectrum obtained by co-adding the various {\it FUSE} exposures with a total integration time of $280$~ks was measured to have a mean $S/N \sim 28$ per 20~{\kms} resolution element for $\lambda > 1000$~{\AA} and significantly lower for $\lambda < 1000$~{\AA}. Certain regions of the {\it FUSE} spectrum were affected by contamination from the interstellar absorption. We used the catalog of molecular hydrogen lines from \citet{wakker06} for this sight line to distinguish possible contamination. The spectra were normalized to the level of a continuum determined using the IRAF{\footnote {IRAF is distributed by the National Optical Astronomy Observatories, which are operated by AURA, Inc., under cooperative agreement with NSF}} SFIT procedure.

\section{Properties of the $z = 0.22496$ Absorber}

The $z(\OVI) = 0.22496$ absorption system detected along the H~$1821+643$ sightline is plotted in Figure 1 and the line measurements are listed in Table 1. In addition to {\OVIdblt}~{\AA} lines, the spectra also show absorption from {\CII}~$\lambda 1335$~{\AA}, {\CIII}~$\lambda 977$~{\AA}, {\OIII}~$\lambda 833$~{\AA},{\SiII}~$\lambda 1260$~{\AA}, and {\SiIII}~$\lambda 1207$~{\AA} lines. The {\CIVdblt}~{\AA} lines associated with this system are detected at lower resolution (FWHM $\sim 30$~{\kms}) in the $STIS$ G230M grating spectrum. The {\SiIV}~$\lambda 1394$~{\AA} is a tentative detection, as we discuss later in this section. The {\HI} absorption is seen in multiple Lyman series lines. The wavelength corresponding to the redshifted {\OIII}~$\lambda 833$~{\AA} line is covered by the lower resolution {\it FUSE} data. The line profile is contaminated by {\SiII}~$\lambda 1021$~{\AA} absorption from the Galactic spiral arm (the Outer Arm) at $v_{LSR} \sim -120$~{\kms}. However, the strength of  the unsaturated Galactic {\SiII}~$\lambda 1526$~{\AA} along this sight line suggests that the {\SiII}~$\lambda 1021$~{\AA} overlapping the {\OIII} absorption is a weak feature with an estimated equivalent width of $W \sim 30$~m{\AA} \footnote {The {\SiII}~$\lambda 1021$~{\AA} line is $\sim 13$ times weaker than the {\SiII}~$\lambda 1526$~{\AA} line.}.  A fit to the {\SiII}~$\lambda 1526$~{\AA} high velocity feature yields $N(\SiII)$ and $b(\SiII)$.  Using these line parameters, we show in Figure 1 the expected contamination from Galactic high-velocity {\SiII}~$\lambda 1021$~{\AA} by synthesizing a line profile . Most of the absorption seen at $\lambda \sim 1020.3$~{\AA} is from the {\OIII}~$\lambda 832$~{\AA} feature associated with the $z = 0.22496$ absorber. Where ever possible we use measurements from fitting theoretical Voigt profiles to absorption lines. In certain cases we also quote the column density derived from the apparent optical depth (AOD) method of \citet{savage91}.  In Table 1, we list the apparent column density \citep{savage91} of {\OIII} by integrating over the $-70$ to $58$~{\kms} velocity interval range. For $v > 58$~{\kms}, there is very strong contamination from low velocity Galactic {\SiII}~$\lambda 1021$~{\AA}. The {\OIV}~$\lambda 787$~{\AA} line is heavily affected by interstellar H$_2$ absorption. The region corresponding to the redshifted {\NeVIIIdblt}~{\AA} has low S/N ($\sim 4$ per 10~{\kms} bin) and thus lacks the sensitivity for a relatively weak line detection. We therefore quote the {\NeVIII} column density as a limit based on non-detection at the $3 \sigma$ significance level. The {\NVdblt}~{\AA} doublet lines are also non-detections. 

The complex velocity profile of {\HI} suggests blending of different absorbing components. We use the automated Voigt profile fitting routine described in \citet{churchill03} to decompose the profile and determine the component structure. We assume Gaussian line spread functions corresponding to the resolving powers of the spectrographs. Applying simultaneous fits to the {\Lya} and higher order Lyman series lines yields a multi-component model that is non-unique (see Table 1). The Voigt profile fits point to some of the {\HI} components being kinematically broad, with $b(\HI) > 30$~{\kms}. If thermally broadened, these line widths imply $T > 3 \times 10^4$~K. In Table 1, we also list the {\HI} line parameters obtained by excluding the saturated {\Lya} and {\Lyb} lines from profile fitting. Even though the higher order Lyman lines are unsaturated, the absorption profiles are not sufficiently well resolved for a definitive identification of the {\HI} component structure. 

Multi-component profiles are also clearly seen in {\OVI}, {\SiIII} and {\CIII} absorption. The individual members of the {\OVIdblt}~{\AA} lines show mild differences in the component structure which are evident when we compare the apparent column density ($N_a(v)$) profiles (see Figure 2). The difference could be due to the effect of noise on one of the lines. Since the Voigt profile fitting technique is sensitive to such uncertainties, the {\OVI} component structure is not well ascertained. The  results from fitting the {\OVIdblt}~{\AA} lines simultaneously and separately are listed in Table 1. 

The multi-component structure for an intermediate ion is best determined from the {\SiIII}~$\lambda 1207$~{\AA} line. A Voigt profile fit decomposes the absorption into three distinct components with narrow line widths (see Table 1). There is a $> 3 \sigma$ detection of a weak {\SiII}~$\lambda 1260$~{\AA} absorption at $v = 28.6$~{\kms} with $b(\SiII) = 4~{\pm}~1$~{\kms}. This absorbing component is tracing the same gas as the {\SiIII}~$\lambda 1207$ component at $v \sim 31$~{\kms} with $b(\SiIII) = 6~{\pm}~1$~{\kms}. The $b(\SiII)$ and $b(\SiIII)$ suggest $T < 6 \times 10^4$~K and $T < 2.7 \times 10^4$~K in that gas. 

The {\CII}~$\lambda 1335$~{\AA} line is detected corresponding to the $v \sim 3$~{\kms} {\SiIII} component. The blue end of the absorption is close to a gap between adjacent echelle orders, which appears as a emission-like feature in the displayed spectrum (see Figure 1). In the complicated {\CIII} absorption profile, at least four components are clearly evident. The positive velocity wing of the {\CIII} profile is fit using a broad component centered at $v = 24$~{\kms} with $b(\CIII) = 64~{\pm}~3$~{\kms} which implies a temperature of $T \lesssim 3 \times 10^6$~K if the line broadening is purely thermal. The other three components of {\CIII} are relatively narrow and aligned with {\SiIII}. 

In the case of {\SiIV}, a weak absorption is detected at the expected wavelength of the $\lambda 1394$~{\AA} line, coincident in velocity with the $v \sim 31$~{\kms} {\SiIII} component. We measure a rest-frame equivalent width of $W_r = 17.8~{\pm}~5.0$~m{\AA} for this feature. However, the feature appears twice as broad as the {\SiII} and {\SiIII} components at that velocity which likely indicates a contribution from a separate gas phase than those low ions.  The {\SiIV} components corresponding to the other two components of {\SiIII} are very weak with a $< 3 \sigma$ detection significance. Hence, we regard the {\SiIV}~$\lambda 1394$~{\AA} for this system as tentative and list both a measured $N(\SiIV)$ and an integrated 3$\sigma$ upper limit. 

Comparison of $N_a(v)$ profiles of the higher order Lyman series lines (e.g. {\HI}~$\lambda 972$~{\AA}) with {\SiIII} and {\OVI} suggests that the {\HI} components are better represented by {\SiIII} than {\OVI} (see Figure 2). This points to bulk of the {\HI} absorption occurring in the {\SiIII} phase rather than {\OVI}. The {\OVIdblt}~{\AA} lines display broader components than {\CII}, {\SiII}, and {\SiIII}. If thermally broadened, the component line widths of $b(\OVI) \gtrsim 15$~{\kms} suggest $T > 2.2 \times 10^5$~K, which is much higher than the temperature that can be attained through photoionization. The differences in the profile shapes between {\OVI} and {\SiIII} further signifies the multiphase nature of this absorber. Nonetheless, the {\HI}, {\SiIII} and {\OVI} absorption span a similar range in velocity ($|\Delta v| \sim 200$~{\kms}). The centroids of their absorbing components are not significantly displaced from each other. The separate gas phases are thus kinematically coupled, and are probably part of the same absorbing structure. In Sec 5 we describe the physical location of this absorber based on imaging data for this quasar field. In Sec 6.1, we discuss the ionization mechanisms in the gas, using constraints imposed by the measured ionic column density ratios and the kinematics of the absorber. 

\section{Properties of the $z = 0.22638$ Absorber}

The measurements for the $z(\OVI) = 0.22638$ absorber are listed in Table 2. The continuum normalized system plot is displayed in Figure 3.  The absorber is displaced by $\Delta v \sim +350$~{\kms} from the $z = 0.22496$ multiphase system. The {\HI} in this absorber is offset in velocity from the {\OVI} by $v = -50~{\pm}~2$~{\kms}, which is significant for trying to understand the physical nature of this absorption system. The line measurements at the velocity of this offset {\HI} are tabulated in Table 3. 

A simultaneous single component Voigt profile fit to {\OVIdblt}~{\AA} lines yields log~$[N(\OVI)~{\cmsq}] = 13.51~{\pm}~0.04$, $b(\OVI) = 16~{\pm}~2$~{\kms}. The $b(\OVI)$ constrains the gas temperature to $T \leq 2.5 \times 10^5$~K. If the line broadening is dominantly thermal, then the implied temperature is higher than what is plausible in photoionized gas. Due to the significant offset of the {\Lya} from the {\OVI} line, the {\HI} that co-exists with the {\OVI} phase needs to be determined separately. By integrating over a $|\Delta v| = 150$~{\kms} velocity interval centered on the {\OVI} absorption, we derive an upper limit of $W_r(\Lya) < 108$~m{\AA} for {\HI} in the same ionized phase as {\OVI}. This translates into a column density of log~$[N(\HI)~{\cmsq}] < 13.4$. The $150$~{\kms} velocity interval is twice the FWHM of an {\HI} line at a $T = 2.5 \times 10^5$~K (based on the {\OVI} line width) that is purely thermally broadened. A closer comparison of the {\Lya} profile with the {\OVI} shows that the {\HI} absorption quickly recovers to the continuum level at positive velocities from the {\OVI} line centroid. Thus, a more stringent $3 \sigma$ upper limit of $W_r(\Lya) < 25$~m{\AA} and log~$[N(\HI)~{\cmsq}] < 12.7$ can be derived by integrating over the positive velocity interval from $v = 0$ to $v = + 75$~{\kms} and subsequently doubling this measurement. 

At the same velocity as {\OVI} there is no  evidence for absorption from other species. The location of the {\OIII}~$\lambda 833$~{\AA} line in the spectrum shows {\OIV}~$\lambda 788$~{\AA} absorption from a strong metal line absorption system at $z \sim 0.2966$. If the relative elemental abundances in the absorber are comparable to solar, then we do not expect to see {\OIII}~$\lambda 833$~{\AA}, since the {\CIII}~$\lambda 977$~{\AA} line, which is usually $\sim 9$ times stronger, is a non-detection. The wavelength corresponding to the redshifted {\OVI}~$\lambda 788$~{\AA} suffers from Galactic H$_2$ contamination. 

The {\Lya} absorption offset from {\OVI} is spread over a velocity interval of $\Delta v = 190$~{\kms}. Our fitting routine derives a two component Voigt profile fit to the {\Lya} feature. Much of the {\HI} absorption is arising in a broad component at $v = -50~{\pm}~2$~{\kms} with log~$[N(\HI)~{\cmsq}] = 13.50~{\pm}~0.01$ and $b(\HI) = 51~{\pm}~2$~{\kms}. The system is therefore a broad {\Lya} absorber (BLA). The line width of the BLA implies a maximum temperature of $T = 1.6 \times 10^5$~K if the line is thermally broadened. The broad profile can also arise due to kinematic effects in the absorbing gas. 

We also detect a shallow and broad {\CIII}~$\lambda 977$~{\AA} absorption feature coincident in velocity with the BLA to within 1$\sigma$ uncertainty. The rest-frame equivalent width of $W_r(\CIII~977) = 17.3~{\pm}~5.4$~m{\AA}, makes it a detection that is statistically significant at the 3~$\sigma$ level. However, the measurement for such a shallow feature is subject to uncertain continuum placement. A single component Voigt fit shows that the {\CIII}~$\lambda 977$~{\AA} line is aligned in velocity with the centroid of the BLA to within 1$\sigma$ uncertainty (see Table 3). If the BLA and {\CIII} arise in the same gas phase, then $b(\HI) = 51~{\pm}~2$~{\kms} and $b(\CIII) = 28~{\pm}~4$~{\kms} imply a thermal component of $b_{t}(\HI) = 43$~{\kms} and a non-thermal component of $b_{nt} = 28$~{\kms} for the line widths. The thermal line width would give $T = 1.1 \times 10^5$~K for the BLA gas phase which is several times larger than the temperature of gas in photoionization equilibrium. We note that the measured $b(\CIII)$ is not significantly affected by instrumental broadening since the FWHM of the $STIS$ E140M observation is only $\sim 7$~{\kms}.  

The kinematic offset between the {\OVI} and {\HI} and their column density ratios point to the absorption arising in regions with different ionization conditions. The {\CIII} appears as tracing the same ionized gas as the BLA, and thus can provide some constraint for the temperature and metallicity in that phase. In Sec 6.2, we consider plausible ionization scenarios for this absorber and evaluate their predictions against observation. 

\section{Galaxies Associated With The Two Absorbers}

Ground based imaging observations of the H~$1821+643$ field have succeeded in identifying several galaxies along this sight line at the absorber redshifts. \citet{tripp98a, tripp00} report the detection of two galaxies with spectroscopic redshifts of $z = 0.2256$ and $z = 0.2265$ in a survey centered on this quasar field. The galaxies are at angular distances of $32^{\prime\prime}.3$ and $114^{\prime\prime}.6$ from the line of sight to the quasar. These correspond to projected linear separations of $\sim 116~h_{71}^{-1}$~kpc and $\sim 412~h_{71}^{-1}$~kpc at $z = 0.2256$ and $z = 0.2265$ \footnote{assuming H$_o = 71$~{\kms}~Mpc$^{-1}$, $\Omega_m = 0.27$ and $\Omega_{\Lambda} = 0.73$ \citep{wright06}.} respectively. For brevity, we will refer to these galaxies as G and W in our discussion. Both galaxies are coincident in redshift with the absorption systems at $z = 0.22496$ and $z = 0.22638$. A more recent survey of the H~$1821+643$ field with the Gemini North telescope has revealed four additional galaxies near this redshift (T. M. Tripp, private communication) but at several times larger impact parameters from the absorber. The quasar sight line is therefore intercepting a group of galaxies at $z \sim 0.226$. 

Savage {\etal}(1998) derive comparable luminosities of $L_B \sim 2 L^*_B$ for the galaxies G and W because of their similar apparent magnitudes of $g = 19.7$ and $g = 19.8$ respectively. The redshift measurements indicate a systemic velocity difference of $|\Delta v| = 158$~{\kms} between the $z = 0.22496$ absorber and galaxy G, and a $|\Delta v| = 377$~{\kms} with galaxy W. The $z = 0.22638$ absorber has a redshift velocity difference of $|\Delta v| = 191$~{\kms} with respect to G and $|\Delta v| = 30$~{\kms} with respect to W. Galaxy G is at a substantially smaller impact parameter from these absorbers compared to W. Based on the relative proximity, Savage {\etal} (1998) attributed the $z = 0.22496$ and the $z = 0.22638$ absorption systems to gas in the extended halo of galaxy G. Such an inference is consistent with the statistical results from past absorber-galaxy surveys \citep[e.g.][]{lanzetta95}, and also with the more recent survey by \citet{wakker09}, where they find the {\OVI} absorbers having an origin within $\sim 500$~kpc of $ > 0.25L^*$ galaxies, with some substantial fraction ($\sim 40$\%) associated with the halos of $> L^*$ galaxies. 

In Figure 4, we display a section of the WIYN $R$ -  band image of the H~$1821+643$ field published by Savage {\etal}(1998) indicating the position of galaxy G with respect to the sight line. The galaxy [RA(2000) = 18:21:54.5,  DEC(2000) = +64:20:09.0] is located at $18^{\prime\prime}.2$ west and $26^{\prime\prime}.7$ south of the quasar, and has an angular radius of $6^{\prime\prime}.6$ that corresponds to $\sim 24$~kpc. The extended source within $\sim 4^{\prime\prime}$ from G is identified as a more distant galaxy at a redshift of $z = 0.2986$ (Schneider {\etal}1992). In Figure 4 we display the only available spectrum of galaxy G from Schneider {\etal}(1992), obtained at a resolution of $\sim 25$~{\AA}. The strong H$\alpha$ + [{\NII}]~$\lambda\lambda 6548, 6584$~{\AA} emission feature was used to determine the redshift of galaxy G to a precision of 0.0005. The relatively narrow width of the Balmer emission feature and the absence of strong [{\OIII}]~$\lambda\lambda 4959, 5007$~{\AA} emission lines indicate that galaxy G does not harbor an active galactic nucleus or an exceptionally high star-formation rate. We note that the width of H$\alpha$ feature in itself is not a strong indicator of high star formation rate (Knapen \& James 2009). Based on careful analyses of the imaging and spectroscopic data, Savage {\etal}(1998) conclude galaxy G to be luminous spiral of morphology Sbc-Sc, consistent with the strength of the H$\alpha$ emission line feature \citep{kennicutt92a, kennicutt92b}. 

From the integrated H$\alpha$ luminosity, we estimate the star-formation rate in galaxy G using the relation, SFR (M$_\odot$ yr$^{-1}$) = $7.9 \times 10^{-42}$ L(H$\alpha$) (ergs s$^{-1}$) from \citet{kennicutt98}. This expression is a revision of Kennicutt (1983a) calibration based on updated stellar models and IMF. The unresolved H$\alpha$ + [$\NII$] emission feature at $\lambda \sim 8050$~{\AA} in the redshifted spectrum of galaxy G (see Figure 4) has an observed equivalent width of $42$~{\AA} \citep{schneider92}. We correct for the [$\NII$] contamination using the standard flux ratio H$\alpha$/[$\NII$] = 2.3 \citep{kennicutt92a}, and also account for a 1 mag internal extinction, typical for the relevant redshift \citep{tresse98}. The L(H$\alpha) \sim 4.3 \times 10^{41}$~ergs s$^{-1}$ thus obtained yields a SFR $\sim 3.4$~M$_\odot$~yr$^{-1}$, which is less compared to nearby starburst galaxies such as M~82 \citep[$\sim 10$~M$_\odot$ yr$^{-1}$, ][]{oconnell78}.  The extinction in H$\alpha$ can be inferred directly by estimating the observed Balmer decrement in the spectrum of galaxy G. However, we are restricted in this approach due to the lack of reliable information on the equivalent width of the weak H$\beta$ emission seen at $\lambda \sim 5955$~{\AA}. Very approximate measures from Figure 4 imply a  (H$\alpha$ + [\NII])/H$\beta$ flux ratio of $\sim 10$. Adapting this flux ratio and using equation (1) in \citet{tresse98} for extinction, we determine the interstellar reddening as $A_v \sim 2.3$~mag. The H$\alpha$ luminosity corrected for this reddening is L(H$\alpha$) $\sim 1.1 \times 10^{42}$~ergs s$^{-1}$, corresponding to a SFR $\sim 8.3$~M$_\odot$ yr$^{-1}$. This is comparable to the SFR in certain nearby actively star forming systems. However we emphasize that the above estimation is at best only an upper limit on the rate of star formation particularly since galaxy G's spectrum is devoid of strong [{\OIII}] and other nebular emission which are characteristic of active star formation. 

If the absorbers are residing in the halo of galaxy G, it is relevant to consider the galaxy as a source of ionizing radiation. The ionization fractions of {\HI}, and low ions such as {\CII} and {\SiII} can be influenced by a galaxy's radiation field. On the other hand, for the ionization fraction of {\OVI} to be altered, photons with energy $E \geq 113.9$~eV are required. The radiative intensity of O and B stars at such high energies is very low due to the significant opacity from He$^+$ ionization edge at $E \geq 54.4$~eV. The {\OVI} ionization levels will remain unaffected by the photons escaping from galaxy G, unless it contains an active nucleus which is inconsistent with its observed integrated spectrum. In Figure 5, we show an estimate of the ionizing spectrum from galaxy G at a distance of 100~kpc. The spectrum was calculated by adapting the model for the radiation escaping from the Milky Way described in \citet{fox05}, and scaling it to the luminosity of galaxy G. This model assumes an escape fraction of 6\% for the hard photons with $\lambda < 912$~{\AA} and 22\% for soft UV photons in the range 912 - 2460~{\AA}.  The galaxy's ionizing flux is compared with the isotropic extragalactic background field of \citet{haardt01}, which has contributions from quasars as well as star-forming galaxies. The surface flux of photons \footnote{The surface flux of hydrogen ionizing photons is calculated as $\Phi = \int_{\nu_o}^{\infty} (F_\nu/h\nu)~d\nu$, where $\nu_o$ is the frequency corresponding to 13.6 eV. In the previous expression $F_\nu = 4\pi J_\nu$, where $J_\nu$ is the specific intensity in units of ergs s$^{-1}$ cm$^{-2}$ Hz$^{-1}$ str$^{-1}$.} with energy $E \geq 13.6$~eV from the extragalactic radiation field at $z = 0.225$ and intercepted by the cloud surface is ($\Phi = 6.3 \times 10^4$~{\cmsq}~s$^{-1}$) found to be a factor of $\sim 6$ stronger than the ionizing flux from the galaxy at a distance of 100 kpc. The surface flux of non-hydrogen ionizing photons from the extragalactic background ($\Phi = 1.8 \times 10^8$~{\cmsq}~s$^{-1}$) is also substantially larger compared to the contribution from the galaxy ($\Phi = 6.7 \times 10^5$~{\cmsq}~s$^{-1}$). 

We note that the above estimation is only an approximation. The true flux of the radiation field from galaxy G at the location of the absorber will depend on several unknown factors including the orientation of the galaxy's disk with respect to the absorbers, the distribution of  O \& B stars, white dwarfs and soft X-ray sources within the galaxy contributing towards the ionizing flux, and the escape fraction of those UV photons into the halo. However, given the 100~kpc separation between the absorber and the galaxy, the ionization of the gas is likely to be dominated by the extragalactic radiation field. Hence, in the photoionization analysis of the absorbers, we use only the \citet{haardt01} model for the radiation field. We discuss in Sec 6.1 the extent to which the results would vary if the estimated number density of ionizing photons ($n_\gamma$ ) from the galaxy is also included. 

\section{Ionization Conditions in The Two Absorbers}

In the following subsections, we evaluate the predictions made by photoionization and collisional ionization models using constraints set by the observed ionic column densities and line widths. Our goal is to differentiate the dominant ionization mechanism in the separate gas phases thereby gaining insight into the physical nature of the absorbers. 

\subsection{$z = 0.22496$ Absorber}

The low ions ({\CII}, {\SiII}), intermediate ions ({\SiIII}, {\CIII}) and high ions ({\OVI}, {\CIV}) can be used to evaluate the ionization of the gas in this absorber. Photoionization calculations were performed using Cloudy [ver.C08.00, \citet{ferland98}] with the initial assumption of solar elemental abundance ratios \footnote{The solar elemental abundances incorporates recent revisions and are as follows; (C/H)$_\odot$ = -3.61 dex, (O/H)$_\odot$ = -3.34 dex, (Si/H)$_\odot$ = -4.46 dex \citep{allende02, holweger01}}. The ionizing background was modeled after the \citet{haardt01} extragalactic radiation field which incorporates UV photons from quasars and star forming galaxies. The component structure is best determined for {\SiIII}~$\lambda 1207$~{\AA} (see Sec 3) and hence we use it to optimize our ionization models. At the same velocity as the $v \sim 31$~{\kms} component, there is a detection of weak {\SiII}~$\lambda 1260$~{\AA}. The {\SiII} and {\SiIII} likely exist in the same gas phase, as explained in Sec 3. Using the $N(\SiII)/N(\SiIII) \sim 0.2$, we determine the ionization parameter for this component as log~$U = -3.0$ \footnote{Ionization parameter is defined as the ratio of the number density of photons with E $ \geq 13.6$~eV to the hydrogen density, $U = n_{\gamma}/n_{\H}$}. The ionization parameter constraint is evident from the photoionization curves displayed in Figure 6. The {\HI} component closest in velocity to the {\SiIII} component has an $N(\HI) = 10^{15.26}$~{\cmsq}. Assuming that all of this {\HI} is associated with the same gas phase as {\SiIII}, we derive a silicon abundance of [Si/H] = -0.62 dex. Synthetic absorption profiles based on the column densities derived from the photoionization model with log~$U = -3.0$, [Z/H] = -0.62 dex are shown in Figure 7. The predicted {\CII}~$\lambda 1335$~{\AA} and {\CIII}~$\lambda 977$~{\AA} for the $v = 31$~{\kms} component are also consistent with the data, which suggests a carbon abundance of $\sim -0.6$~dex (see Figure 6). For this component, the above model yields a total hydrogen column density of $N(\H) = 3.9 \times 10^{17}$~{\cmsq}, a volume density of $n_{\H} = 2.5 \times 10^{-3}$~{\cc}, a photoionization equilibrium temperature of $T = 1.3 \times 10^4$~K, and a gas pressure of $p/K = 71$~K~{\cc}. The density and total hydrogen column density imply a path length of $L = 0.05$~kpc for the gas cloud probed by this absorption component. {\it We find that the predicted {\OVI} column density in this photoionized gas phase (see Figure 6) is $\sim 5$~dex smaller than the observed value}. 

There is uncertainty in the Si and C abundances, dominated by systematic uncertainties associated with the hydrogen profile modeling and the assumptions inherent in Cloudy photoionization calculations. The $N({\HI})$ used to constrain the metallicity is based on the simplistic assumption that the profile structure, as seen in the higher order Lyman lines, is well represented by a two component Gaussian (see Figure 7). The error values given by the profile fitting procedure do not adequately account for the uncertainty in the {\HI} component structure. The true $N(\HI)$ associated with the $v = 31$~{\kms} {\SiIII}  is possibly within 0.2~dex of the fit measurement. Also, the uncertain shape of the ionizing UV background from quasars results in a $\sim 0.3$~dex uncertainty in the metallicity predicted by Cloudy \citep{aracil06}. Considering these systematics,  we estimate the uncertainty in the Si and C abundances to be $0.4$~dex.

For the $v \sim 3$~{\kms} {\SiIII} component, there is weak {\CII}~$\lambda 1335$~{\AA} absorption detected. The corresponding {\CIII} is saturated such that $N(\CII)/N(\CIII) < 0.3$. The ratio provides a lower limit of log~$U > -3.0$, corresponding to a density upper limit of $n_{\H} < 1.9 \times 10^{-3}$~{\cc} (see Figure 6), comparable to the density determined for the 31~{\kms} component. This ionization parameter is also consistent with $N(\SiIII)/N(\SiIV) \gtrsim 0.4$. We note that the {\CII} is a very weak feature, and the {\SiIV} only a tentative detection (see Sec 3). Hence the respective ionic column density ratios are, at best, limits. For the absorbing component at $v \sim -39$~{\kms} the constraint on log~$U$ is weak due to the non-detection of {\CII} or {\SiII}. Assuming solar relative elemental abundance for carbon and silicon, the $N(\CIII)/N(\SiIII) \sim 20$ is recovered at log~$U \sim -2.2$ (see Figure 6). Due to its complicated velocity profile, the {\HI} corresponding to the  $v \sim 3$~{\kms} and $v \sim -39$~{\kms} components are uncertain and hence an abundance estimation is not meaningful. 


The strength of {\OVI} relative to {\CIV}, and the absence of {\SiIV}  and {\NV} absorption indicates that this separate phase is heavily ionized. The lower resolution ($R = \lambda/\Delta \lambda \sim 10,000$, FWHM $\sim 30$~{\kms}) G230M grating spectrum does not resolve the component structure in the {\CIVdblt}~{\AA} lines. To a first approximation, we can determine the ionization conditions in the {\OVI} gas phase by comparing the integrated apparent optical depth column densities of {\OVI} and {\CIV}. The ratio $N_a(\CIV)/N_a(\OVI) \sim 0.2$ ratio is recovered through photoionization at an ionization parameter of log~$U \sim -0.8$ which corresponds to $n_{\H} = 1.1 \times 10^{-5}$~{\cc}. If we assume that this {\OVI} absorbing gas has the same abundances as the {\SiIII} in the $31$~{\kms} component, then a log~$U = -0.8$ model yields $N(\H) = 7.9 \times 10^{18}$~{\cmsq}, $n_{\H} = 1.1 \times 10^{-5}$~{\cc}, $T = 3.3 \times 10^5$~K, $p/K = 8.2$~{\cc} K, and a path length of $\sim 230$~kpc. Such a large path length and low pressure is unrealistic given the close kinematic coupling between {\OVI} and the lower ionization gas traced by {\SiIII}, {\CIII}, {\SiII}, and {\CII} which has higher pressure and much smaller physical size. Increasing the metallicity in the {\OVI} gas to solar would still yield a path length of $\sim 60$~kpc, which is still $\sim 3$ orders of magnitude bigger than the size of the lower ionization gas clouds. 

A more realistic scenario is the production of {\OVI} through collisional processes in a higher temperature gas. Collisional ionization equilibrium (CIE) models predict the measured $N_a(\CIV)/N_a(\OVI) \sim 0.2$ at $T \sim 2 \times 10^5$~K. Such a CIE temperature is consistent with the absence of {\SiIV} and {\NV}, and also $N_a(\CIII)/N_a(\OVI) \gtrsim 0.5$ (assuming solar abundance ratios). However, gas in the temperature range of $T \sim (1 - 5) \times 10^5$~K undergoes rapid radiative cooling and will be displaced from a state of thermal equilibrium. Non-equilibrium collisional ionization processes thus become important for the production of {\OVI} in this absorber.  

In multiphase gaseous environments such as the Galactic high velocity clouds (HVCs) and the local interstellar medium (LISM), nonequilibrium collisional ionization models have been particularly successful in explaining the origin of {\OVI} \citep{sembach03, fox04, savage06}. In the next section, we compare the $z = 0.22496$ absorption system to Galactic high velocity gas where the origin of {\OVI} is better understood. The proximity of the $z = 0.22496$ absorber to a luminous galaxy (see Sec 5) makes such a comparison relevant. 

\subsubsection{The $z = 0.22496$ Absorber as an Extragalactic Analog of Galactic HVC}

Galactic HVCs detected in high ions are multi-phase systems with a mix of warm ($T \sim 10^4$~K) gas traced by low/intermediate ions and a higher ionization gas phase at a higher temperature ($T \sim 10^5$~K) traced primarily by {\OVI} and {\CIV} \citep{sembach00, sembach03, fox04, fox05, ganguly05}. The higher temperature is usually evident from the line width of  {\OVI} components which are, on average, broader than the components seen in low/intermediate ions. The $N(\OVI)/N(\HI)$ in Galactic HVC samples span a wide range ($\sim 10^{-6} - 10^{-2}$), even though the observed $N(\OVI)$ is usually within 1 dex of $\sim 10^{13}$~{\cmsq}, suggestive of entirely different ionization mechanisms operating in the {\HI} and {\OVI} gas phases \citep{fox05}. In HVCs, photoionization under the influence of an extragalactic radiation field or a radiation field escaping from the Galaxy is found to be sufficient to explain the observed strength of low/intermediate ions, whereas the $N({\OVI})$ (and to some extent {\CIV}) is inconsistent with a similar origin \citep[e.g. ][]{fox04, fox05}. Such overall properties of the highly ionized Galactic HVCs are remarkably similar to what we find for the $z = 0.22496$ absorber (see Sec 6.1). The photoionized components in Galactic HVCs have densities [$n_{\H} \sim (1 - 10) \times 10^{-3}$~{\cc}], and  physical sizes  [$L \sim (20 - 500)$~pc] which are in the range of what we derive for the {\CII}, {\SiII}, {\SiIII} phase in the $z = 0.22496$ absorber. Collisional ionization models for Galactic HVCs, with temperatures typically of $T \sim 1.7 \times 10^5$~K well reproduce the observed ionic ratios such as $N_a(\CIII)/N_a(\OVI)$ \citep{fox06}. This is also consistent with the temperature for the {\OVI} phase given by the component line widths in the $z = 0.22496$ absorber.

Based on careful analysis of the close kinematic association of {\OVI} absorption with low/intermediate metal ions, and the column density ratio predictions provided by various ionization models, several authors have proposed that the Galactic high velocity {\OVI} is most likely produced at interface layers between {\it warm} gas clouds and a {\it hot} exterior medium \citep[e.g. ][]{sembach03, fox04, fox05}. In the halo, the {\it hot} exterior medium corresponds to the diffuse [$n_{\H} \sim (1 - 10) \times 10^{-5}$~{\cc}] coronal plasma with $T \sim (1 - 3) \times 10^6$~K extending to galactocentric distances of $R > 70$~kpc. The steep temperature gradient between the {\it warm} photoionized HVC phase and the {\it hot} exterior gas gives way to heat transport through electron collisions resulting in transition temperatures $T \sim (1-6) \times 10^5$~K at the interface layers \citep{savage06}. In the interface layers equilibrium ionization conditions are unlikely and  the ionization state of the gas cannot be described using a single temperature.  Among the various nonequilibrium processes, conductive interface models \citep{boehringer87, borkowski90} have been particularly successful in reproducing the range of high ion column density ratios, and also the velocity offsets between the absorption from the {\it warm} phase and the interface gas \citep{fox04, fox05}. The high ion ratios of $N_a(\CIV)/N_a(\OVI) \sim 0.2$, $N_a(\NV)/N_a(\OVI) < 0.1$, $N_a(\SiIV)/N_a(\OVI) < 0.04$ which we measure for the $z = 0.22496$ absorber are fully consistent with a conductive interface model \citep[see Figure 12 of][]{fox05}. A single conductive interface layer is predicted by the models to yield a $N(\OVI) \sim 10^{13}$~{\cmsq}. This value is $\sim 1.5$~dex lower than the observed integrated {\OVI} column density. In the context of Galactic HVCs, \citet{sembach03} have shown that higher values of integrated {\OVI} column densities can be explained using multiple interface layers formed at the boundaries of the several small-scale fragments within the clouds. The multi-component profile of {\OVI} in the $z = 0.22496$ absorption system is consistent with such a scenario. Also, the $T \sim (1 - 6) \times 10^5$~K expected for the transition temperature plasma agrees with the $b(\OVI) \gtrsim 15$~{\kms} in the individual components of the absorption. 

A distinct possibility thus exists for the $z = 0.22496$ absorber to be a multiphase high velocity cloud system embedded within the halo of galaxy G, the galaxy closest to the absorption system, at a projected distance of $\sim 116~h_{71}^{-1}$~kpc from the sight line. The boundary between the {\it warm} photoionized high velocity gas and the {\it hot} halo enveloping the galaxy can give rise to transition temperature gas where {\OVI} (and much of {\CIV}) gets produced through collisional processes. 

The physical conditions in the {\it warm} photoionized gas of the $z = 0.22496$ high velocity absorber can be used to impose limits on the properties of the {\it hot} halo of galaxy G. In the context of the Milky Way, HVC structures are understood to be confined by the thermal pressure of the halo gas \citep{murali00, stanimirovic02}. Assuming an isothermal halo of $T \sim 2 \times 10^6$~K, pressure confinement predicts the density of the gaseous halo of galaxy G to be $n_{\H} \sim 10^{-5} $~{\cc} at $100$~kpc. Detecting HVC structures thus extends a direct observational constraint on the gas density in the diffuse outer halo of galaxies. By assuming a spherical halo of radius $R_h = 200$~kpc and uniform density, we obtain a rough estimate for galaxy G halo gas mass as $\sim 6 \times 10^9$~M$_\odot$. 

\subsection{$z = 0.22638$ Absorber}

In this absorber, we have fewer column density measurements to ascertain the ionization in the separate phases. The offset in velocity of $v \sim -50$~{\kms} between {\OVI} and {\HI} suggests that the two ions, though spatially related, are tracing distinct gas phases \citep{tripp00, tripp08}. The broad and shallow {\Lya} profile suggests that the absorber is a BLA (see Sec. 4). In the following sections, we treat the ionization conditions in the {\OVI} (Sec 6.2.1, and 6.2.2) and BLA (Sec 6.2.3) gas phases separately. 

\subsubsection{Can the {\OVI} be Photoionized ?}

The photoionization code Cloudy (ver C08.00) was used to investigate the possibility of {\OVI} tracing photoionized gas, with further constraints imposed by the non-detection of low/intermediate ions and {\CIV}.  The ionizing radiation field was modeled after \citet{haardt01}. For a given combination of metallicity and ionization parameter, the models were optimized to converge on the measured $N(\OVI)$. The photoionization results are tabulated in Table 4. In the {\OVI} phase, the {\HI} to {\OVI} column density ratio is constrained to $N(\HI)/N(\OVI) < 10^{-1}$ (see Table 2). Figure 8 is a visualization of the {\HI} line profiles predicted by the various photoionization models. As these synthetic profiles show, even for a very high ionization parameter limit such as log~$U < -0.1$, at [Z/H] = -0.5 dex, {\HI} is overproduced compared to data, evident from the excess absorption at positive velocities with respect to the {\OVI} line centroid. Moreover, for a log~$U \sim -0.1$, the implied density is so low that it results in a very large path length ($d \sim 0.4$~Mpc) for the absorbing region. The Hubble flow broadening from a such a large path length is going to be a factor of $\sim 1.5$ larger than the measured $b(\OVI) = 16~{\pm}~2$~{\kms}. 

A photoionization solution simultaneously satisfying the path length constraint as well as the $N(\HI)/N(\OVI)$, $N(\CIII)/N(\OVI)$ and $N(\CIV)/N(\OVI)$ limits is possible only at [Z/H] $\gtrsim 0$ dex. Even at solar metallicity, the ionization parameter in the {\OVI} gas phase is limited to a narrow range of $-0.8 < $~ log~$U < -0.1$, corresponding to a very low gas density of $n_{\H} \sim 10^{-6}$~{\cc}. Even in these models, the size of the {\OVI} absorbing region is predicted to be $d \gtrsim 10$~kpc (see Table 4). A 10~kpc path length seems incompatible with the relatively symmetric and narrow absorption profile of {\OVI}. The models estimate an equilibrium temperature of $T \sim 3 \times 10^4$~K which suggests that $> 90$\% of the {\OVI} line width must be due to non-thermal effects. 

In summary, a photoionization origin is unrealistic for the {\OVI} in the $z = 0.22638$ absorber. The very low densities predicted for photoionized medium ($n_{\H} \sim 10^{-6}$~{\cc}) in order to not overproduce {\CIII}, {\CIV} and {\HI} from the same phase as {\OVI}, results in large absorber sizes for [Z/H] $\leq 0$~dex.  A large path length for the absorbing region is inconsistent with the symmetric and relatively narrow line profile of {\OVI}, which shows no direct evidence of subcomponent structure.

\subsubsection {Can the {\OVI} be Collisionally Ionized?}

A second possibility is the production of {\OVI} through the collision of electrons with ions in a medium with $T \gtrsim 10^5$~K. \citet{thom08} show that $N(\OVI)/N(\HI) > 1$  is possible in collisionally ionized gas even when the metal abundance is low. In the $z = 0.22638$ {\OVI} absorber, the temperature implied by the {\OVI} line width is $T \leq 3.1 \times 10^5$~K (i.e., $T \leq 10^{5.5}$~K), which is close to the temperature at which {\OVI} ionization fraction peaks under CIE. If line broadening is predominantly thermal, then the temperature upper limit is sufficiently large to produce detectable amounts of {\OVI}. The temperature upper limit is also consistent with the non-detection of {\CIII}, {\CIV} and {\NeVIII} ions. The non-detection of {\CIII} and {\CIV} at the same velocity as {\OVI} can be used to place a lower limit on the gas temperature. For $T > 10^5$~K, the {\CIII} ionization fraction rapidly declines. In CIE, the $N(\OVI) > N(\CIII)$ and $N(\OVI) > N(\CIV)$ for $T \gtrsim 2.0 \times 10^5$~K. Therefore we can limit the temperature in the {\OVI} gas phase to a narrow range of $(2.0 - 3.1) \times 10^5$~K (i.e., $10^{5.3}$~K $- 10^{5.5}$~K). In that range of temperatures, most of the oxygen will be in the O$^{3+}$ state. However, the wavelength corresponding to the redshifted ${\OIV}~\lambda 788$~{\AA} line is contaminated by very strong Galactic H$_2$ absorption, and thus  we lack this crucial additional constraint. 

For any given combination of $N(\HI)$ and [Z/H], there is a unique temperature at which the observed value of $N(\OVI)$ is recovered, assuming CIE conditions. For example, as shown in Figure 9, at $N(\HI) = 10^{12.6}$~{\cmsq} and [Z/H] = -0.9 dex, the measured $N(\OVI)$ is reproduced by the CIE model at a temperature of $T = 3.2 \times 10^5$~K, which also satisfies the $N({\CIV}) < 10^{12.8}$~{\cmsq} limit. The $N(\CIII)$ predicted by the CIE model is marginally consistent with the 3$\sigma$ upper limit. Since the $N(\HI) = 10^{12.6}$~{\cmsq} is only an upper limit for {\HI} at the same velocity as {\OVI} (see Table 2), the metallicity estimation is indefinite. Also, the CIE model assumes a (C/O) relative elemental abundance ratio of solar, which is also not well constrained by the data. 

The $T = (2.0 - 3.1) \times 10^5$~K predicted for the {\OVI} phase is in the temperature range where radiative cooling is efficient. Departures from CIE are expected due to rapid metal line cooling. In Figure 7, we show the {\OVI}, {\CIII} and {\CIV} column densities at [Z/H] = -0.9 dex for an isochorically cooling gas as computed by \citet{gnat07}. The ionization fractions of {\OVI} as a function of temperature does not show significant departure from CIE at -0.9 dex metallicity. Both the CIE and non-CIE predicitions for the {\CIII} and {\CIV} column densities are also consistent with their respective upper limits. From the available column density information alone it is difficult to conclude whether the gas is in ionization equilibrium or not. 

\subsubsection {The BLA Gas Phase}

The {\HI} at $v = -53$~{\kms} from the {\OVIdblt} lines is a BLA (see Sec 4). The temperature implied by the $b(\HI) = 51~{\pm}~2$~{\kms} is $T \sim 1.6 \times 10^5$~K, assuming pure thermal broadening. There is a formal 3$\sigma$ detection of {\CIII}~$\lambda 977$~{\AA} line, but no {\OVI}, at the velocity of BLA. If the {\HI} and {\CIII} are tracing the same gas, then $b(\HI)$ and $b(\CIII)$ imply $T = 1.1 \times 10^5$~K. This value is a factor of $\sim 4$ higher than the temperatures attained through photoionization. Nonetheless, we explore a range of parameter space in metallicity and log~$U$ to evaluate the feasibility of photoionization in the BLA phase. The predictions of the various models are tabulated in Table 5. It is evident that photoionization equilibrium models do not succeed in simultaneously explaining the {\HI}, {\CIII}, {\CIV} and {\OVI} column density measurements and limits. More importantly, the predicted temperatures are inconsistent with the expected value from the {\HI} and {\CIII} line widths. It is possible for the {\CIII}~$\lambda 977$~{\AA} line to be a blend of unresolved components of smaller line widths. However, it is difficult to discriminate such a multicomponent structure in the present data, since the {\CIII} feature is very shallow and the $S/N$ of the data inadequate. 

The temperature given by $b(\HI)$ and $b(\CIII)$ suggest collisional ionization to be a more realistic scenario. At $T = 1.1 \times 10^5$~K, the $N(\CIII)$ predictions made by CIE and non-CIE models differ by $\sim 0.4$~dex. For the given $N(\HI)$ and temperature, the non-CIE models estimate the C abundance to be [C/H] $\sim -1.5$~dex, whereas the CIE models estimate the abundance as $\sim -1.8$~dex. At the temperature given by the line widths, radiative cooling can be rapid leading to significant departures from equilibrium ionization fractions. Hence the non-CIE abundance estimation is preferred. 

In Figure 10, we display the non-CIE ionization curves for {\CIII}, {\CIV} and {\OVI} in an isochorically cooling gas at a chosen metallicity \citep{gnat07}. For comparison, the corresponding CIE predictions are also shown. At [C/H] $=-1.5$~dex and $T = 1.1 \times 10^5$~K, the non-CIE model is able to the reproduce the observed {\CIII} to {\HI} column density ratio. The $f(\HI) = N(\HI)/N(\H) \sim 10^{-5}$ ionization correction given by the non-CIE model results in a total hydrogen column density of $N(\H) \sim 3.2 \times 10^{18}$~{\cmsq},  signifying the BLA to be a substantial baryon reservoir. 

\subsubsection {Physical Nature of the $z = 0.22638$ Absorber}

It is evident from the analysis presented in the foregoing two sections that the {\OVI} and the {\HI} in the $z = 0.22638$ absorber are tracing separate collisionally ionized gas phases. The absorber is situated in the vicinity of a luminous galaxy (see Sec 5), and thus it could potentially be a gasesous structure embedded within the galaxy's extended halo. The kinematic proximity of this system to the higher column density $z = 0.22496$ absorber also leads to this possibility (see Sec 6.1.1). However, the {\OVI} in this absorber might not be produced in a typical {\it warm} - {\it hot} gas interface as in the multiphase Galactic HVCs, since we do not detect a {\it warm} gas phase for this absorber.  At the boundary between the BLA gas and the {\it hot} corona of galaxy G, transition temperature gas would still form due to the temperature gradient. The $T = 1.1 \times 10^5$~K for the BLA gas and the $T \gtrsim 10^6$~K for the {\it hot} corona would result in transition temperatures within the  $T = (2.0 - 3.1) \times 10^5$~K range expected for the {\OVI} gas (see Sec 6.2.2). 

The low carbon abundance of [C/H] $= -1.5$~dex and $T \sim 10^5$~K suggest that the BLA could be a fragmentation within the {\it hot} halo of galaxy G, born out of small-scale density or temperature perturbations. The presence of such condensation clouds in halos are a prediction of certain semi-analytical galaxy formation models \citep{maller04}. In this scenario, the BLA gas must be a young condensation ($\sim 10^5$~yrs) that has not cooled to the typical {\it warm} temperatures observed in HVCs. The {\OVI} could in turn be produced at the turbulent layers formed at the boundary between the condensing cloud at $T \sim 10^5$~K and the {\it hot} exterior medium described as the galaxy's corona. 

Kinematic displacement between the {\OVI} and the {\HI} could arise if the {\OVI} gas is tracing an evaporative flow within the conductive interface layer \citep{boehringer87, borkowski90}.  In the context of the local interstellar medium, \citet{savage06} find that significant velocity offsets ($\Delta v \sim 30$~{\kms}) between {\OVI} absorption and tracers of the {\it warm} gas are reproducible in the evaporative phase of the interface evolution. Such evaporative flows from the {\it warm} to the {\it hot} medium are common during the initial $\sim 2 \times 10^6$~yrs of an interface. We find that the $N(\OVI) \sim 10^{13}$~{\cmsq} predicted for the evaporative stage is consistent with the {\OVI} column density measurement. However, there are two concerns with this possibility. Firstly, theoretical models for a young interface typically predict transition temperatures that are a factor of $\sim 2$ higher than the maximum temperature permitted by $b(\OVI) \sim 16$~{\kms}, although these predictions are dependent on model assumptions. Secondly, the magnitude of the velocity offset depends on how steep the temperature gradient is between the separate gas phases that form the interface layer. For the $z = 0.22638$ system, the difference in temperature between the BLA and the {\it hot} corona may not be sufficient to generate flow velocities which fully account for the measured $v \sim -50$~{\kms} offset between the {\OVI} and {\HI} line centroids. 

Another circumstance that can result in a velocity offset is if the BLA and {\OVI} are tracing separate regions of shock heated gas formed when the condensing cloud is moving through a {\it hot} low density halo \citep{quilis01}. The collisionally ionized {\it post shock} region (gas behind a shock front) can produce {\OVI} at temperatures consistent with the observed line width of {\OVI} \citep{tumlinson05}. The velocity offset between the post shock region and the shock front will depend on the flow velocity and the geometry of the line of sight through the structure. With only one constraint (the offset velocity), it is difficult to explore such models. 

To summarize, the absorber $z = 0.22638$ is a multiphase system. The column densities of {\OVI}, {\CIII} and {\HI} and their line widths suggests that collisional processes control the ionization in the absorber. The BLA is tracing $T \sim 10^5$~K gas that could be part of a  cooling condensing structure within the halo which has not reached temperatures typical of the {\it warm} phase of the HVCs. We lack sufficient information to conclude on any unique physical model for the kinematic offset of $\sim -50$~{\kms} between the {\OVI} and BLA gas phases except that the kinematic offset is broadly consistent with the general predictions of conductive interface models as well as shock heated gas. 

\section{Discussion}

Our interpretation of the astrophysical nature of the two absorbers, and the environments they trace, are meaningful in the light of some recent results connecting {\OVI} absorption systems to galaxies. From a survey of galaxies in quasar fields with known {\OVI} absorbers, \citet{chen09} found that approximately $90$\% (10/11) of the absorbing galaxies in their sample are emission line dominated. Those galaxies that are absorption dominated do not have {\OVI} detected down to an equivalent width limit of $W_r(\OVI~1032) \lesssim 0.03$~{\AA} in the spectrum of the background quasar. Thus {\OVI} seems to be preferentially selecting gas-rich galaxies. The absorbing galaxies were also found to be part of group environments with their morphologies pointing to past events of interactions or satellite accretion. Such interactions were expected to contribute to the $\sim 64$\% covering fraction of {\OVI} estimated from their sample. The star formation rate which we derive from the strength of the H$\alpha$ emission in the spectrum of galaxy G suggests on going star formation, consistent with a gas-rich system (see Sec 5). It extends the possibility of the $z = 0.22496$ and $z = 0.22638$ {\OVI} absorbers tracing gaseous structures which are associated with the extended environments of galaxy G.

\citet{tumlinson05b} have described the properties of two multiphase {\OVI} absorbers along the line of sight to the low-$z$ QSO PG~$1211+143$. A galaxy survey of the surrounding field showed that the two absorbers are situated within $\sim 150h^{-1}$~kpc of $\sim L^*$ galaxies. One of those galaxies is part of a spiral-dominated group. The {\OVI} in both those absorption systems are inconsistent with a photoionization origin. The $b(\OVI)$ suggests gas temperatures of $T \gtrsim 10^5$~K. Ion ratios in the two absorbers resemble the highly ionized Galactic HVCs. \citet{tumlinson05} propose that these absorbers are related to the nearby galaxies by outflows or tidal streams created from interactions with unseen satellite galaxies. It is possible for the $z = 0.22496$ absorber along the H~$1821+643$ sight line to have an analogous origin. The multicomponent, multiphase absorption profile and the abundance ratio [Si/H] $\sim -0.6$~dex in the photoionized gas agrees with a description of the absorber tracing tidal debris embedded in the halo of its nearby galaxy G. An extended geometry like that of the Magellanic stream would result in several conductive layers at regions where the high velocity gas interfaces with the galaxy's {\it hot} corona. The line of sight intercepting such multiple interface layers would explain the significant velocity spread of $\Delta v \sim 200$~{\kms} in {\OVI} and {\HI} absorption as well as the $N_a(\OVI) \sim 10^{14.3}$~{\cmsq}. 

The physical nature of the $z = 0.22638$ system is relevant for other BLAs with metals detected in the same gas phase. BLAs are characterized by large line widths $b \geq 40$~{\kms} and shallow absorption $N(\HI) < 10^{14}$~{\cmsq} \citep{richter04, richter06}. It is unclear whether the large line width is dominated by thermal or non-thermal effects. Thermal broadening would imply that BLAs are large reservoirs of baryons in the low-$z$ IGM. Detecting metals in the same phase as the broad {\Lya} gas will enable an accurate temperature estimation. For the $z = 0.22638$ BLA, the $b(\HI)$ and $b(\CIII)$ imply $T \sim 1.1 \times 10^5$~K. The temperature indicates the gas to be heavily ionized ($f_{\HI} = N(\HI)/N(H) \sim 10^{-5}$) with a substantial baryon column density of $N(\H) \sim 3.2 \times 10^{18}$~{\cmsq}. 

The $z = 0.16339$ BLA absorber with $b(\HI) = 46.3~{\pm}~1.9$~{\kms} detected in the HE~$0226-4110$ spectrum (Lehner {\etal}2006) resembles the $z = 0.22638$ absorber. In the $z = 0.16339$ system, there is a marginal ($2.9\sigma$) {\CIII} detection that appears shallow and broad, with no associated {\OVI}. The lack of {\OVI} detection points to low metallicity in the gas. If the gas is in CIE at $T = 1.3 \times 10^5$~K suggested by the line width of {\HI}, then the $N_a(\CIII)/N_a(\HI) \sim 0.01$ ratio yields a carbon abundance of [C/H] $\sim -1.8$~dex, an ionization correction of $f({\HI}) \sim 9 \times 10^{-6}$ and a $N(\H) \sim 2.5 \times 10^{19}$~{\cmsq}. These values are comparable to the low abundance and the high baryon content for the BLA phase in the $z = 0.22638$ system. Interestingly, the $z = 0.16339$ BLA is also coincident in redshift ($z = 0.1630$) with a relatively bright galaxy ($m_B = 23.74$) at a projected separation of $\sim 226~h_{71}^{-1}$~kpc \citep{chen09}. The $|\Delta v| \sim 117$~{\kms} velocity separation between the absorber and the galaxy favors an origin for the absorption in some high velocity gas associated with the galaxy. The similarities in derived properties between the $z = 0.22638$ and $z = 0.16339$ systems and their proximity to galaxies further emphasizes the likelihood of BLAs with metals in the same gas phase selecting low metallicity high velocity gas systems in external galaxies. The prospect for detecting such extragalactic absorbers is going to be significantly augmented in the near future with the advent of higher sensitivity observations using the Cosmic Origins Spectrograph. 

The $z = 0.22496$ and $z = 0.224638$ absorbers being extragalactic high velocity clouds has implications for our understanding of the nature of other {\OVI} absorption systems. In a sample of 51 {\OVI} systems at $z < 0.5$ identified along 16 sight lines, Tripp {\etal}(2008) found 53\% of the absorbers to be complex multiphase systems with significant velocity offset between the {\OVI} and {\HI} absorbing components. Even among the fraction of absorbers with closely aligned {\OVI} and {\HI} components, the temperature implied by the combined line widths of {\HI} and {\OVI} in 26\% of the cases was $T \gtrsim 4 \times 10^4$~K. This temperature lower limit is inconsistent with photoionized gas. In such absorbers, nonequilibrium collisional ionization process is a distinct possibility for the production of {\OVI}. The {\OVI} could be tracing transition temperature plasma in high velocity gas clouds embedded within a galaxy halo. The fact that many {\OVI} absorbers are clustered around galaxies further supports this possibility \citep{wakker09, chen09}. 

\section{Summary}

We have analyzed the properties of $z = 0.22496$ and $z = 0.22638$ multiphase absorption systems detected along the sight line to the quasar H~$1821+643$. The measurements are based on a combination of ultraviolet spectra from HST/{\it STIS} and  {\it FUSE}. We also use the existing imaging observations of this quasar field from Savage {\etal}(1998) for information on galaxies in the vicinity of the absorber. The significant results from our analysis are summarized as follows : 

(1) The $z = 0.22496$ and $z = 0.22638$ {\OVI} absorbers are likely situated within the extended halo of a $\sim 2L^*$ Sbc-Sc galaxy within an impact parameter of $\sim 116~h_{71}^{-1}$~kpc from the line of sight.  The absorbers are at 158~{\kms} and 191~{\kms} with respect to the systemic redshift of this galaxy. Also coincident in redshift are four other galaxies, at much larger impact parameters, revealing a group environment. 

(2) The $z = 0.22496$ system is a multiphase, multicomponent absorber detected in {\HI}, {\CII}, {\SiII}, {\CIII}, {\SiIII}, {\OIII}, {\CIV}, and {\OVI}. {\SiIV} is tentatively detected. Comparison of apparent column density profiles show that the {\HI} profile (particularly in higher order Lyman series lines) is better represented by the narrow multicomponent {\SiIII} absorption than by {\OVI}. The {\OVI} absorbing components are broader than the components detected in {\SiIII} and {\SiII}. The kinematics of these components and their line widths suggest that the low/intermediate ions such as {\SiII}, {\CII} and {\SiIII} are tracing a lower ionization and cooler gas phase than the phase traced by {\OVI}. This lower ionization gas produces the bulk of the {\HI} absorption. 

(3) The ionic column density ratios and the measured line widths of {\CII}, {\SiII}, and {\SiIII} in the $z = 0.22496$ absorber are consistent with an origin in a medium that is predominantly photoionized. In this photoionized phase, the $N(\SiII)/N(\SiIII) = 0.2$ in one of the absorbing components yields a density of $n_{\H} = 2.5 \times 10^{-3}$~{\cc}, a total hydrogen column density of $N(\H) = 3.9 \times 10^{17}$~{\cmsq}, a photoionization equilibrium temperature of $T = 1.3 \times 10^4$~K, a gas pressure of $p/K \sim 71$~{\cc} K and a path length of $L \sim 51$~pc for the absorbing region. The Si and C elemental abundances in this component are estimated as $-0.6~{\pm}~0.4$~dex.

(4) We find the $z = 0.22496$ absorber to be analogous to Milky Way highly ionized HVCs detected in low, intermediate and high ions. Similar to the ionization conditions in highly ionized Galactic HVCs, the low/intermediate ions and bulk of the {\HI} are consistent with having an origin in a {\it warm} ($T < 5 \times 10^{4}$~K ) photoionized gas phase, where as the high ions (predominantly {\OVI}) are more likely produced in a collisionally ionized transition temperature plasma ($T \sim (2 - 3) \times 10^5$~K). The transition temperature gas is most likely formed at the interface layers between the {\it hot} ($T \geq 10^6$~K) coronal halo of the galaxy and the {\it warm} HVC cloud embedded within it. 

(5) If the photoionzed gas in the $z = 0.22496$ high velocity cloud is confined by the hot halo, then assuming a isothermal halo of $T \sim 2 \times 10^6$~K, we derive a density of $n_{\H} \sim 10^{-5} $~{\cc} at $100$~kpc for the hot halo. 

(6) The $z = 0.22638$ system is a multiphase absorber with an offset of $v \sim -50$~{\kms} between the {\OVI} and {\HI} absorption. At the velocity of {\OVI}, we measure $N_a(\HI)/N_a(\OVI) < 0.2$. The {\OVI} line width of $b(\OVI) = 16~{\pm}~2$~{\kms} implies $T \sim 3 \times 10^5$~K. The absence of {\HI} and {\CIII} absorption at the velocity of {\OVI} is inconsistent with a photoionization origin for {\OVI}. 

(7) The {\HI} line width in the $z = 0.22638$ system implies that it is a BLA absorber. Coincident in velocity with the BLA, weak {\CIII} is detected, but no {\OVI}. The $b(\HI) = 51~{\pm}~2$~{\kms} and $b(\CIII) = 28~{\pm}~4$~{\kms} implies $T = 1.1 \times 10^5$~K in this gas phase. At this temperature departures from CIE ionization fractions are significant for certain ions due to metal line cooling. Non-CIE models estimate a carbon elemental abundance of -1.5~dex in the BLA based on the $N(\CIII)/N(\HI) \sim 0.1$ constraint. The BLA represents heavily  ionized gas ($N(\HI)/N(\H) \sim 10^{-5}$) with a total hydrogen column density of $N(\H) \sim 3.2 \times 10^{18}$~{\cmsq} signifying a large baryon reservoir. This low metallicity BLA could be metal poor material accreted from the group environment or one of the many cooling fragmentations born out of small-scale thermal instabilities within the {\it hot} galactic halo as proposed by Maller \& Bullock (2004). It could  also be tracing the shock-front of a cloud moving through the halo where the gas is compressed and heated to $T \sim 10^5$~K. 

(8)  The collisionally ionized {\OVI} in the $z = 0.22638$ absorber could be produced in post-shock heated gas trailing behind a low metallicity cloud moving at high velocity (traced by {\HI} and {\CIII}) through the halo of galaxy G. Alternatively, it could also be the evaporative gas of a relatively young ($ < 10^6$~yrs) conductive interface between the $T \sim 10^5$~K BLA and the {\it hot} corona of the galaxy. The kinematic offset between the BLA and the {\OVI} are roughly consistent with the general predictions of both these models. 

\vspace{0.6in}

The presentation of results in this paper has benefited from discussions with Todd M. Tripp. We thank him and an anonymous referee for the several valuable comments on this paper. We are also grateful to Orly Gnat for providing computational data for the non-CIE models. We also thank Gary Ferland and collaborators for making the Cloudy photoionization code publicly available. This research is supported by the NASA {\it Cosmic Origins Spectrograph} Program through a sub-contract to the University of Wisconsin, Madison  from the University of Colorado, Boulder. B.P.W acknowledges support from NASA grant NNX-07AH426. 

\pagebreak


\begin{sidewaysfigure}
\begin{center}
\includegraphics[scale=0.7]{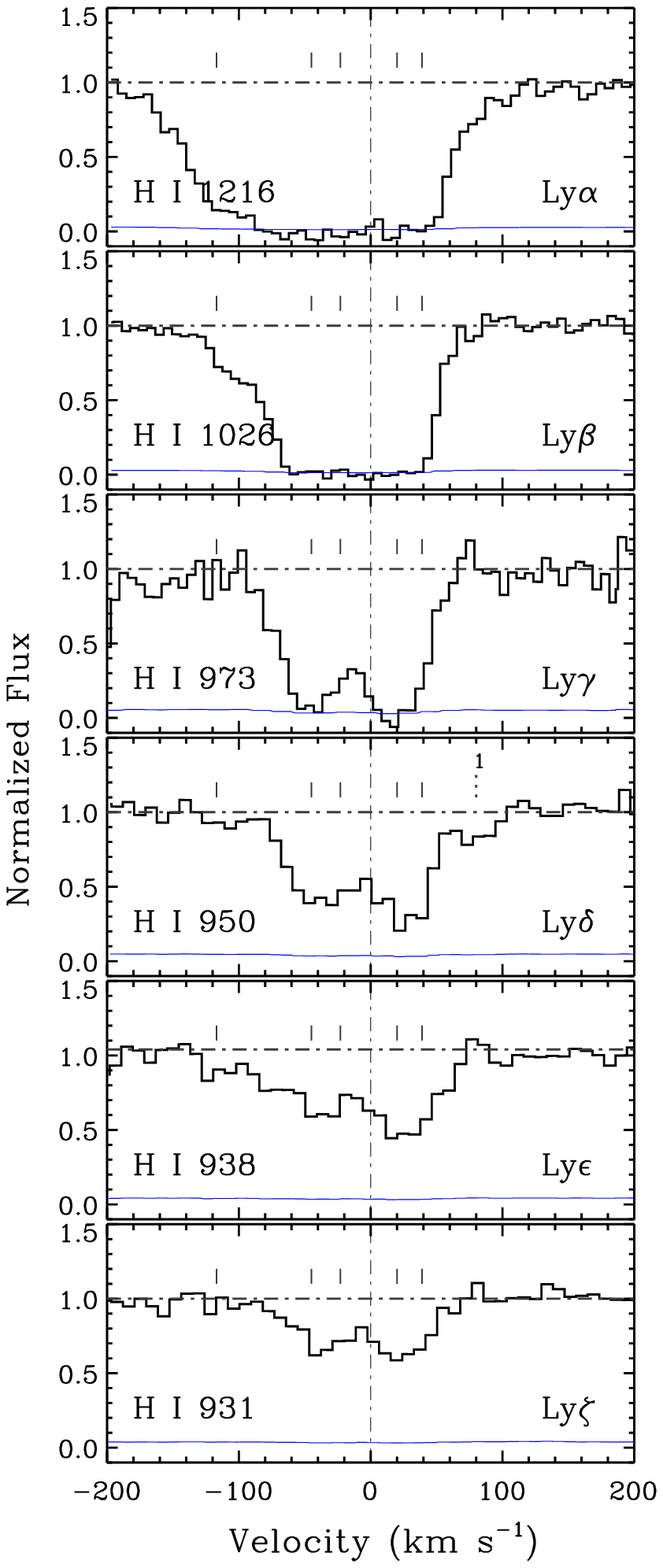}
\hspace{0.3in}\includegraphics[scale=0.7]{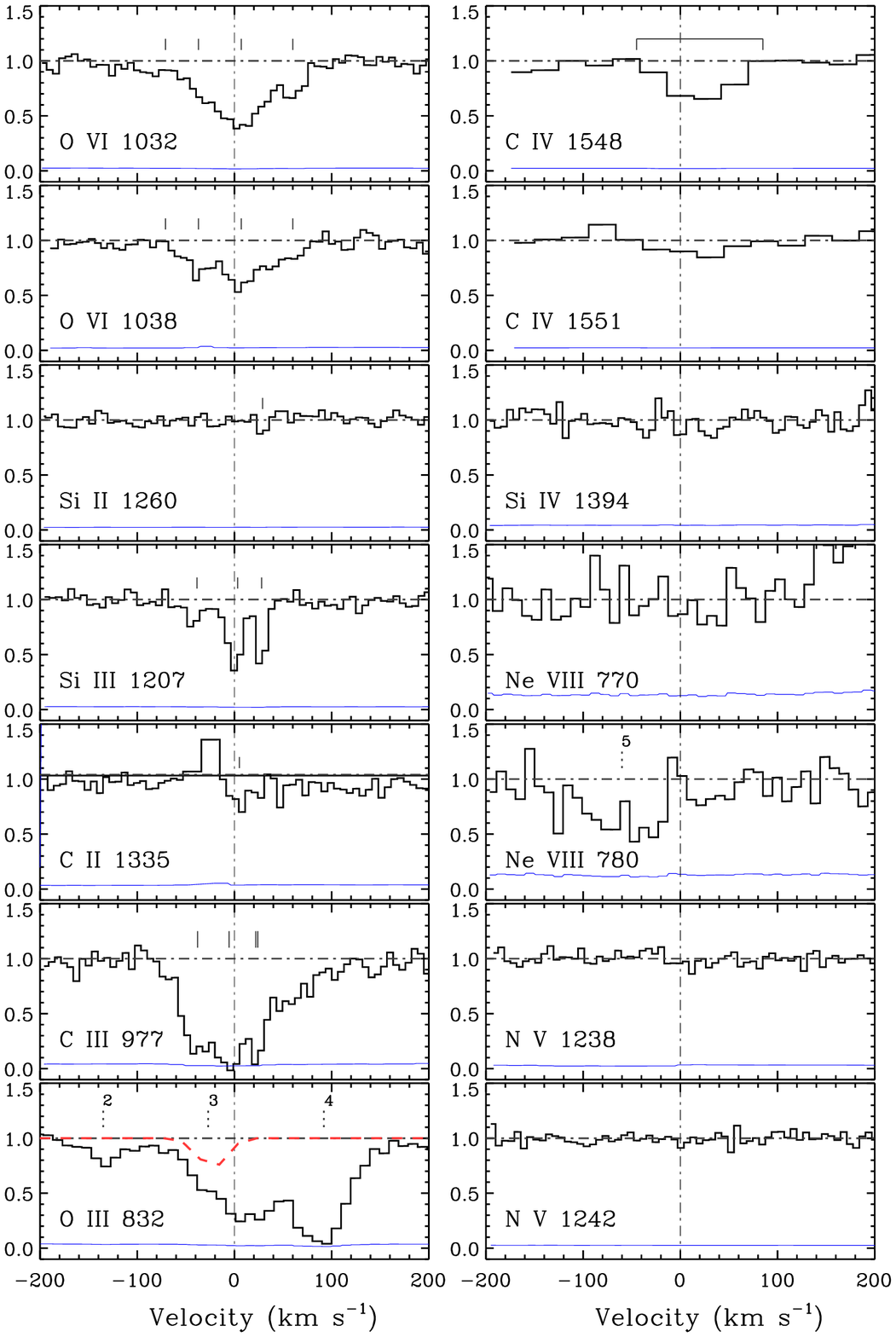}
\end{center}
\protect
\caption{System plot for the $z(\OVI) = 0.22496$ absorber in the continuum normalized spectrum of H~$1821+643$. The centroids of the individual absorbing components for each line obtained from Voigt profile fitting are labeled with vertical tick marks. The {\CIVdblt}~{\AA} and the {\SiIV}~$\lambda 1403$~{\AA} lines are low resolution $STIS$ G230M grating data. The {\OIII}~$\lambda 833$~{\AA} line and the {\NeVIIIdblt}~{\AA} lines are from a low resolution {\it FUSE} spectrum. The {\OIII} line is contaminated by Galactic {\SiII}. The estimated contamination is shown using the synthetic line profile ({\it red dotted line}) superimposed on the spectrum. The velocity interval over which the {\CIV}~$\lambda 1548$~{\AA} was integrated to obtain the apparent column density is marked. The measurements are listed in Table 1. Lines that are not part of the system are labeled and identified as follows (1) {\OVI}~$\lambda 1038$~{\AA} from a metal line system at $z = 0.1214$ \citep{tripp01}, confirmed by the presence of the {\OVI}~$\lambda 1032$~{\AA} line at $\lambda \sim 1157$~{\AA}, {\Lya} and {\Lyb}, (2), (3) and (4) Galactic {\SiII} lines (5) H$_2$ line.}
\label{fig:1}
\end{sidewaysfigure}

\begin{figure*}
\begin{center}
\epsscale{1.0}
\plotone{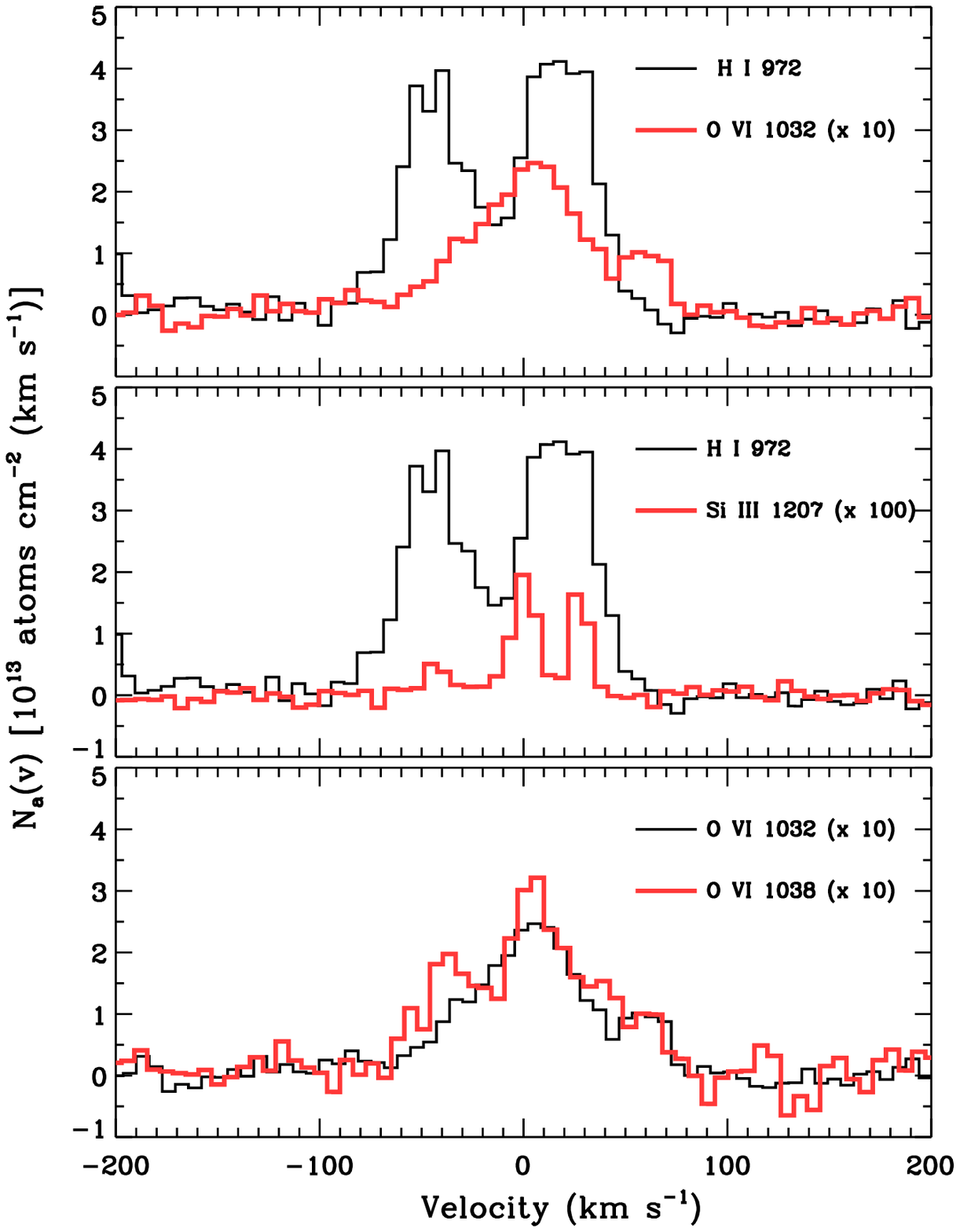}
\end{center}
\protect
\caption{Comparison of apparent column density [$N_a(v)$] profiles of {\HI}~$\lambda 972$~{\AA} with {\OVI}~$\lambda 1032$~{\AA} ({\it top panel}) and with {\SiIII}~$\lambda 1207$~{\AA} ({\it middle panel}) for the $z = 0.22496$ system. Even though the {\HI}, {\OVI} and {\SiIII} absorptions span approximately the same velocity interval, the component structure seen in the {\HI} closely follows the component structure seen in {\SiIII} rather than {\OVI}. This is indicative of the absorber having multiple gas phases with higher ionization conditions (or higher metallicity) in the {\OVI} gas phase compared to {\SiIII}. The {\SiIII} presumably traces gas of lower ionization where bulk of the {\HI} absorption is also arising. Comparison of apparent column density profiles of the individual members of the {\OVI} doublet are shown in the {\it bottom panel}. The profiles show slight variations in the component structure which could be due to the effect of noise on one of the lines.}
\label{fig:1}
\end{figure*}

\begin{figure*}
\begin{center}
\epsscale{0.9}
\plotone{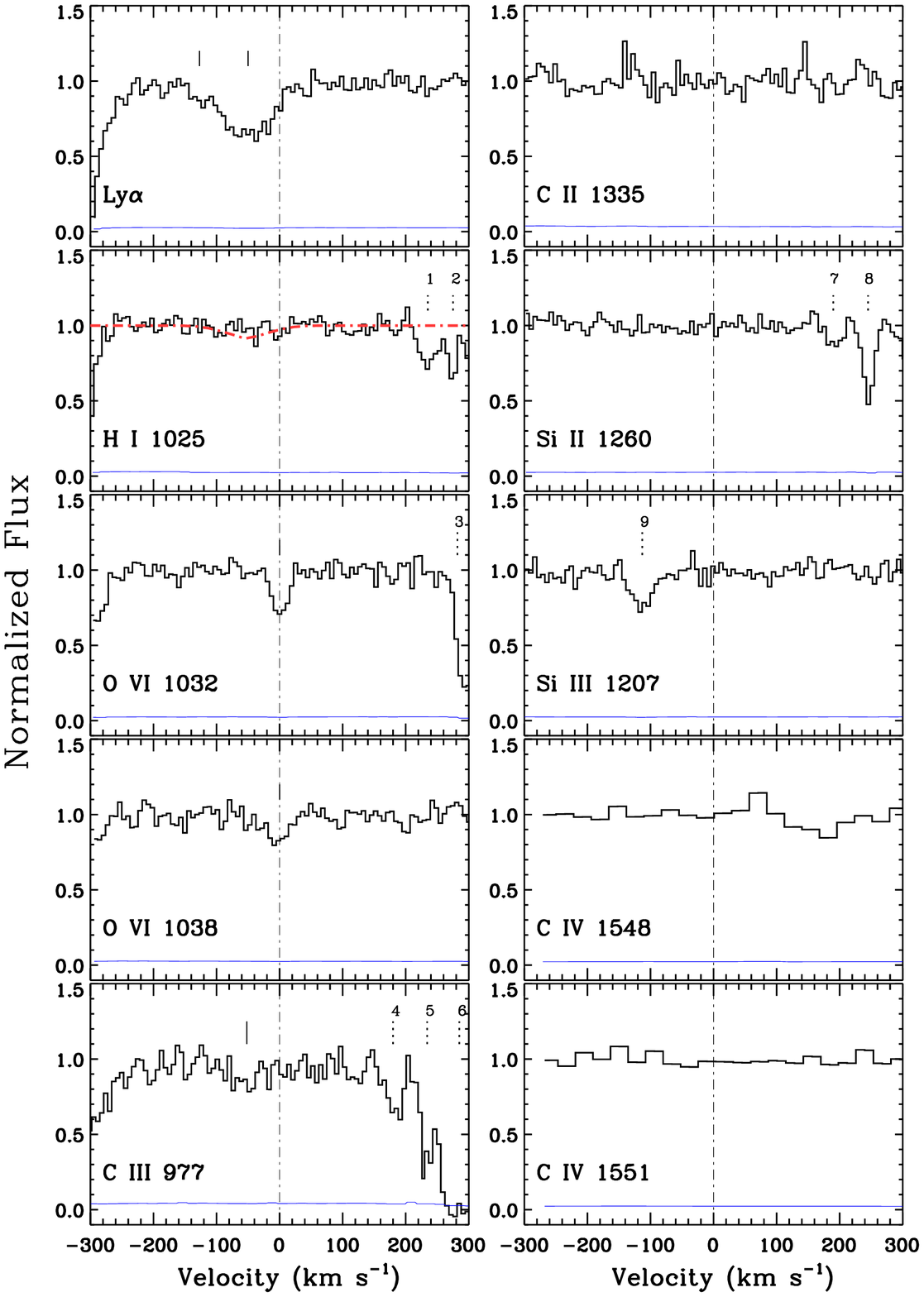}
\end{center}
\protect
\caption{Rest-frame velocity centered system plot for the $z(\OVI) = 0.22638$ absorber in the continuum normalized spectrum of H~$1821+643$.  The centroids of the individual absorbing components for each line derived from Voigt profile fitting are labeled with vertical tick marks. The {\CIVdblt}~{\AA} profiles are lower resolution ($R = \lambda/\Delta \lambda \sim 10,000$) $STIS$ G230M grating data. The centroids of {\Lya} and {\CIII}~$\lambda 977$~{\AA} are offset from the {\OVI}~$\lambda 1032$~{\AA} line centroid by $|\Delta v| \sim -50$~{\kms}. The predicted {\Lyb} absorption profile corresponding to the two components seen in {\Lya} is synthesized and over-plotted ({\it red dash-dot line}) on the relevant panel. The feature is weak and shallow and hence not detected. The measurements are listed in Table 2 and 3. Lines that are not part of the system are labeled and are identified as follows (1) {\OVI}~$\lambda 1038$~{\AA} from a metal line system at $z = 0.21337$ confirmed by the detection of {\OVI}~$\lambda 1032$~{\AA}, {\Lya}, {\Lyb} and {\Lyg} at the same velocity, (2) Galactic {\SII}~$\lambda 1259$~{\AA} at $v_{LSR} \sim -15$~{\kms}, (3) intrinsic {\CIII}~$\lambda 977$~{\AA}; (4), (5) \& (6) Galactic N I~$\lambda 1199$ at $v_{LSR} \sim -136, -86, -31$~{\kms} respectively; (7) \& (8) Galactic {\CIV}~$\lambda 1548$~{\AA} at $v_{LSR} \sim -275, -209$~{\kms} respectively, (9) {\Lya} at $z = 0.21674$.}
\label{fig:1}
\end{figure*}

\begin{sidewaysfigure}
\begin{center}
\epsscale{1.1}
\plotone{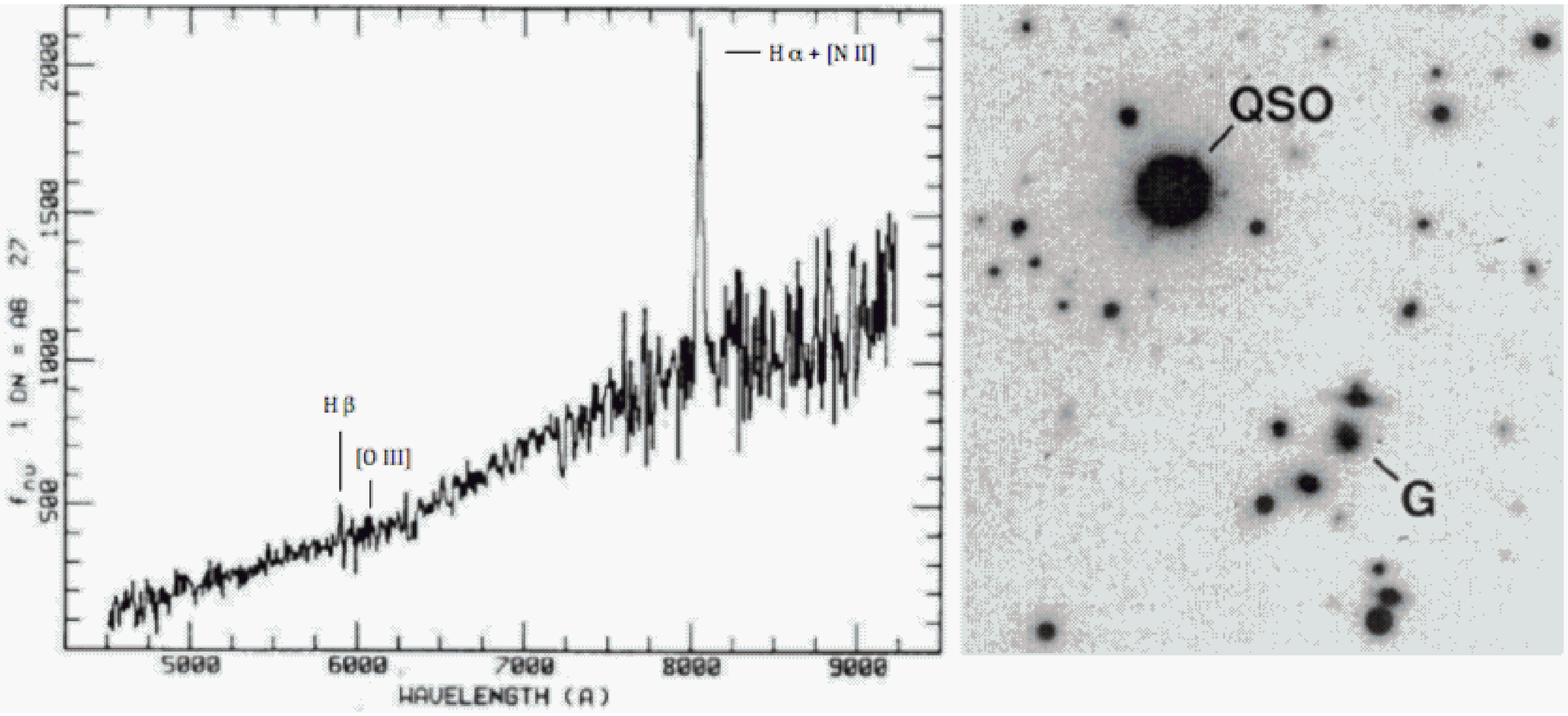}
\end{center}
\protect
\caption{The {\it right} panel shows a section of the 10 minute integration WIYN $R$ - band image of the H~$1821+643$ quasar field of size $1^{\prime}.7 \times 1^{\prime}.8$ with North to the top and East to the left. The image is taken from Savage {\etal}(1998). The {\it left} panel is $\sim 25$~{\AA} resolution spectrum of galaxy G published by Schneider {\etal}(1992). The Y-axis is flux density in units of $5.75 \times 10^{-31}$~ergs {\cmsq} s$^{-1}$ Hz$^{-1}$ corresponding to AB magnitude of 27. The redshift of the galaxy is determined as $z = 0.2265$ from the strong emission feature at $\lambda \sim 8125$~{\AA} which is a blend of H $\alpha$ and [{\NII}]. The weak H$\beta$ and the expected location of the [{\OIII}]~$\lambda 5007$~{\AA} emission features are marked. The galaxy G is at an impact parameter of $\rho \sim 116~h_{71}^{-1}$~kpc from the line of sight. The extended source within $\sim 4^{\prime\prime}$ from G is identified as a more distant galaxy at a redshift of $z = 0.2986$ (Schneider {\etal}1992).}
\label{fig:1}
\end{sidewaysfigure}

\begin{figure*}
\begin{center}
\epsscale{0.9}
\plotone{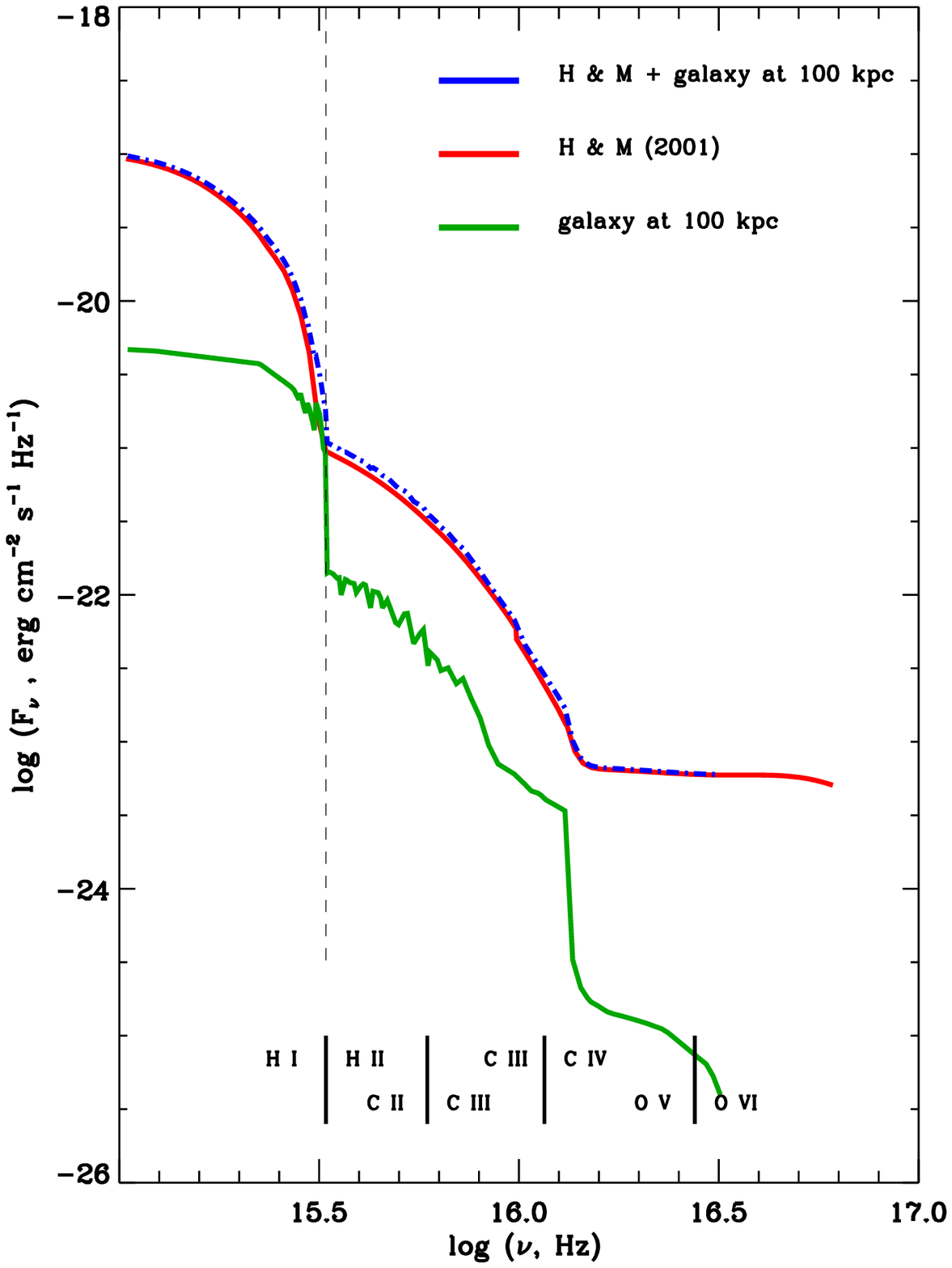}
\end{center}
\protect
\caption{A comparison of the radiation fields relevant for the ionization of $z = 0.22496$ and $z = 0.22638$ absorbers. The extragalactic ionizing background ({\it red}) is obtained from Haardt \& Madau (2001), which includes contribution from quasars, AGNs and photons escaping from young star-forming galaxies.  An estimate of the galaxy's spectrum based on the calculations by Bland-Hawthorn \& Maloney (1999) is shown in {\it blue}. The flux of escaping radiation from the galaxy has been scaled down to a distance of 100~kpc corresponding to the impact parameter of the galaxy G from the location of the absorbers. The combined ionizing flux from the galaxy and the extragalactic background radiation is shown in {\it blue} dash-dot line.The ionization edges of some of the key ions are marked. It is evident that the contribution from the galaxy is insignificant compared to the extragalactic ionizing radiation at $z = 0.225$.}
\label{fig:1}
\end{figure*}

\begin{figure*}
\begin{center}
\epsscale{1.0}
\plotone{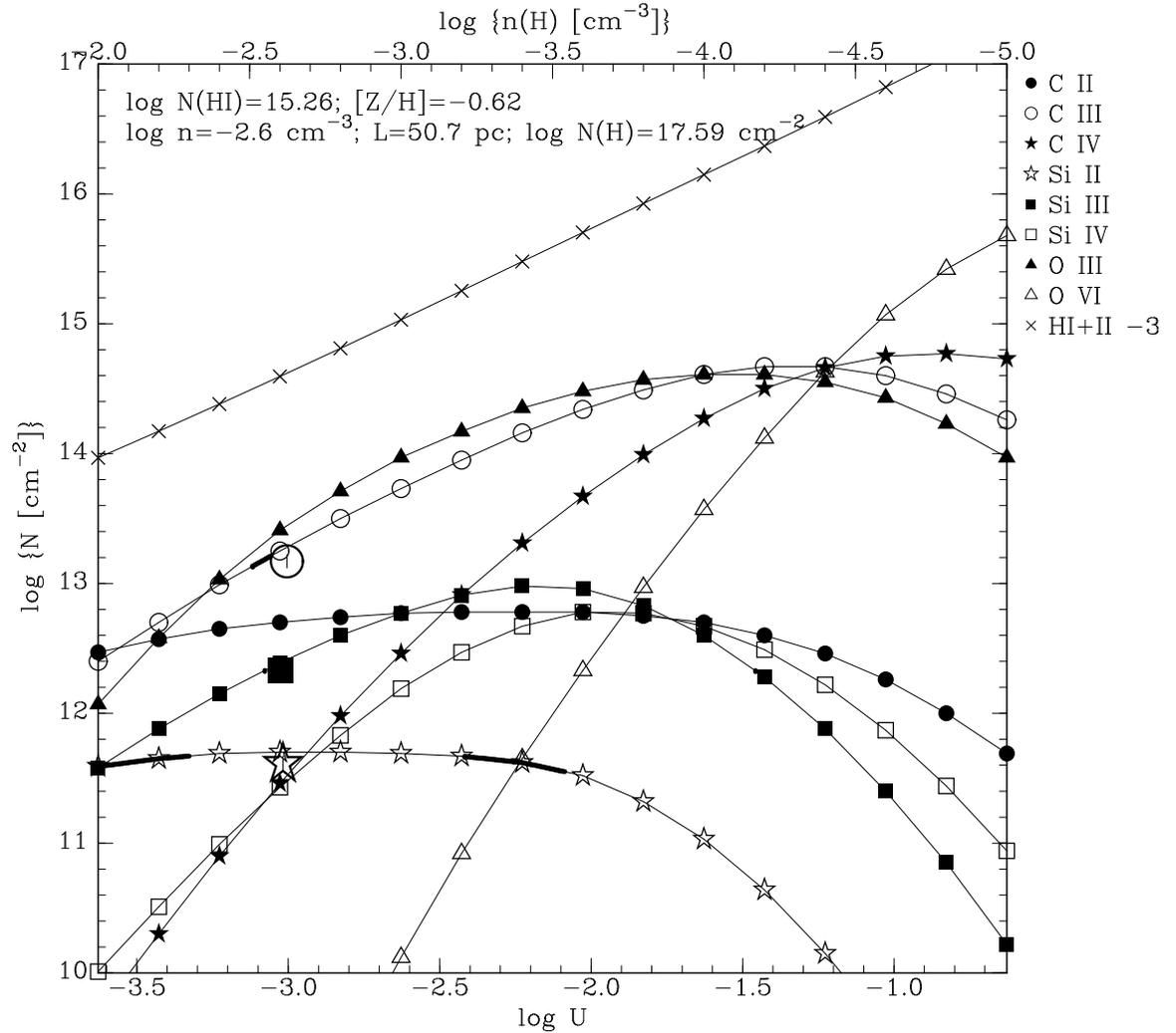}
\end{center}
\protect
\caption{\small Expected column densities from photoionization models for the $v \sim 31$~{\kms} component seen in {\SiIII} and {\SiII} of the $z = 0.22496$ absorber in the spectrum of H~$1821+643$. The $N(\HI) = 10^{15.02}$~{\cmsq} is the {\HI} that is coincident in velocity with this component (see Table 1). The ionizing background is from the Haardt \& Maadu (2001) model for the relevant redshift and includes ionizing photons from quasars and star forming galaxies. The model that best fits the measured $N(\SiII)/N(\SiIII)$ yields an ionization parameter of log~$U = -2.6$ corresponding to a total {\H} density of $n_{\H} = 2.5 \times 10^{-3}$~{\cc}, a total {\H} column density of $N(\H) = 3.9 \times 10^{17}$~{\cmsq} and thus a physical size of $L = 0.05$~kpc for the absorbing region. The measured $N(\SiII)$, $N(\SiIII)$, and $N(\CIII)$ are marked using respective symbols of bigger size. The predicted high ion ({\OVI}, {\CIV}) column densities from this photoionized phase are several dex smaller than the observed values.}
\label{fig:1}
\end{figure*}

\begin{figure*}
\begin{center}
\epsscale{0.8}
\plotone{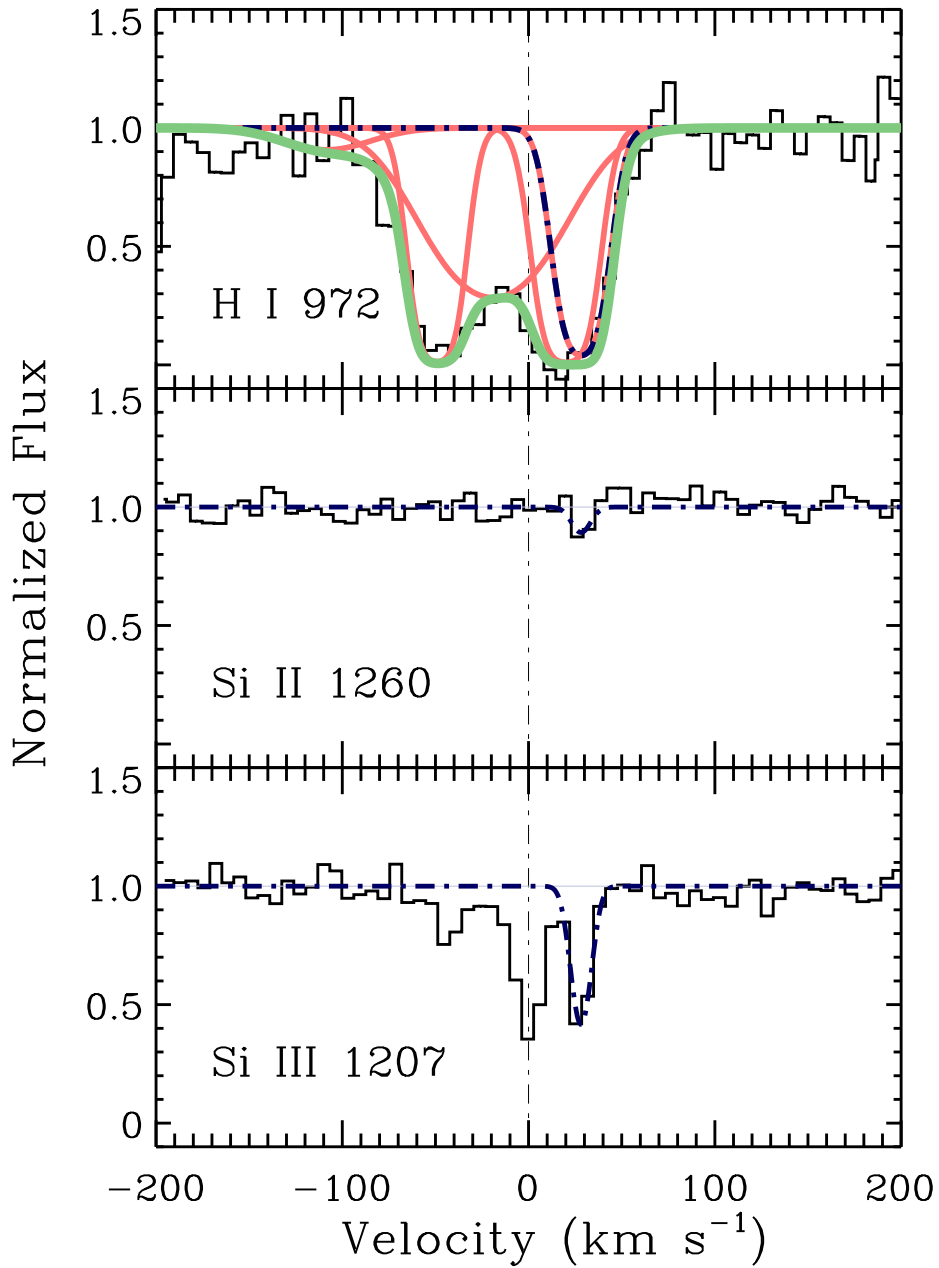}
\end{center}
\protect
\caption{\small Synthetic absorption profiles for $v \sim 31$~{\kms} {\SiIII} and the corresponding {\SiII} based on column densities predicted by a photoionization model with log~$U = -3.0$ is over-plotted ({\it blue}, dash-dotted) on top of the data for the $z = 0.22496$ absorber. The photoionization models were optimized to recover the observed $N(\SiIII)$. The $v \sim 31$~{\kms} {\SiIII} component provides the best constraint for density in the photoionized phase of this absorber due to the weak {\SiII} detection at the same velocity. The {\SiII} to {\SiIII} column density ratio gives the density in the photoionized phase as $n_{\H} = 2.5 \times 10^{-3}$~{\cc}.The individual Voigt profile components for {\HI} are plotted in {\it red}, and the combined Voigt profile model in {\it green}. An estimate of the silicon abundance in the $v \sim 31$~{\kms} photoionized gas is derived by using the {\HI} column density from the two component profile fit shown for the higher order Lyman lines. The {\HI} component structure is not well determined (see Sec 6.1).}
\label{fig:1}
\end{figure*}

\begin{sidewaysfigure}
\begin{center}
\epsscale{1.0}
\plottwo{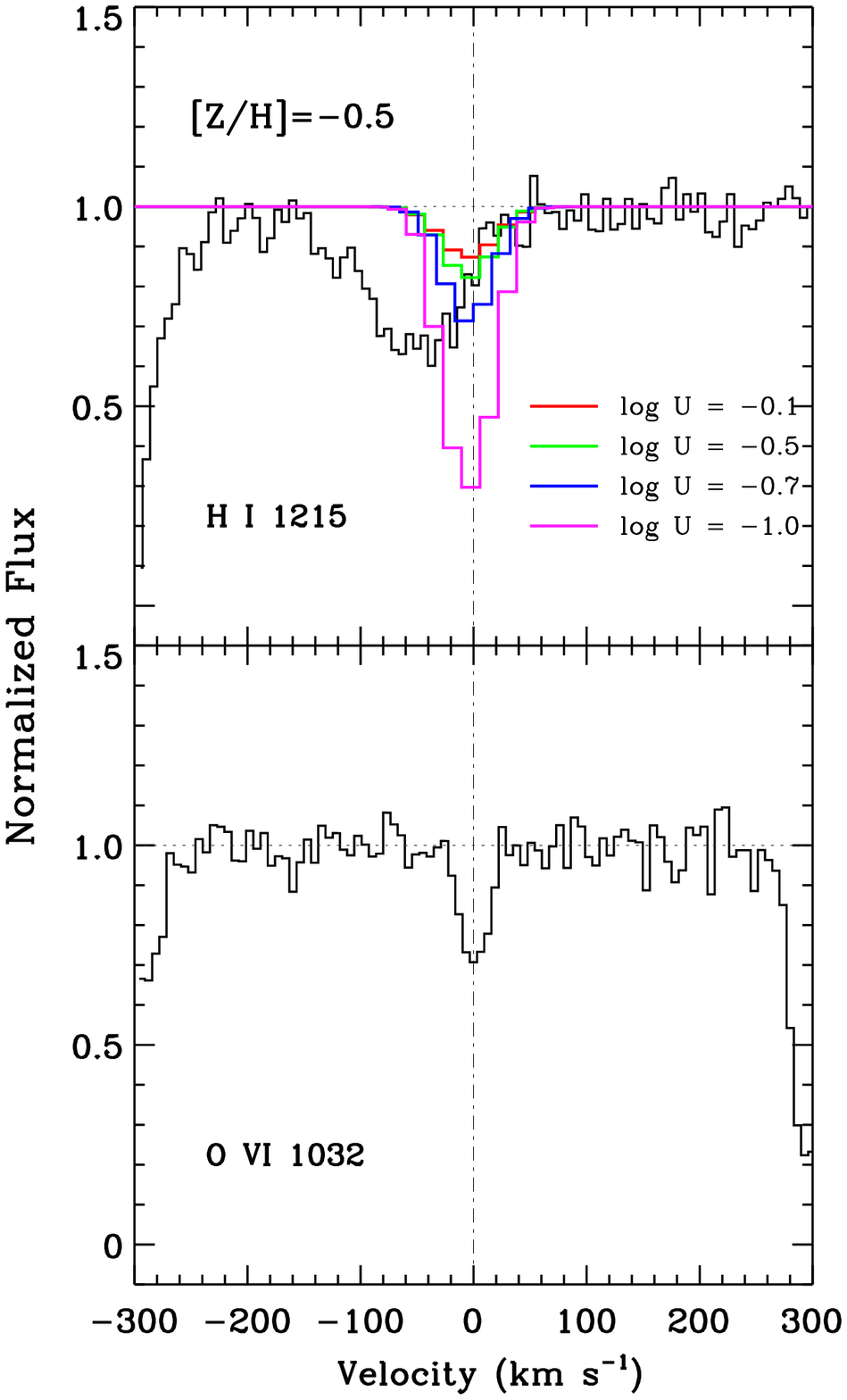}{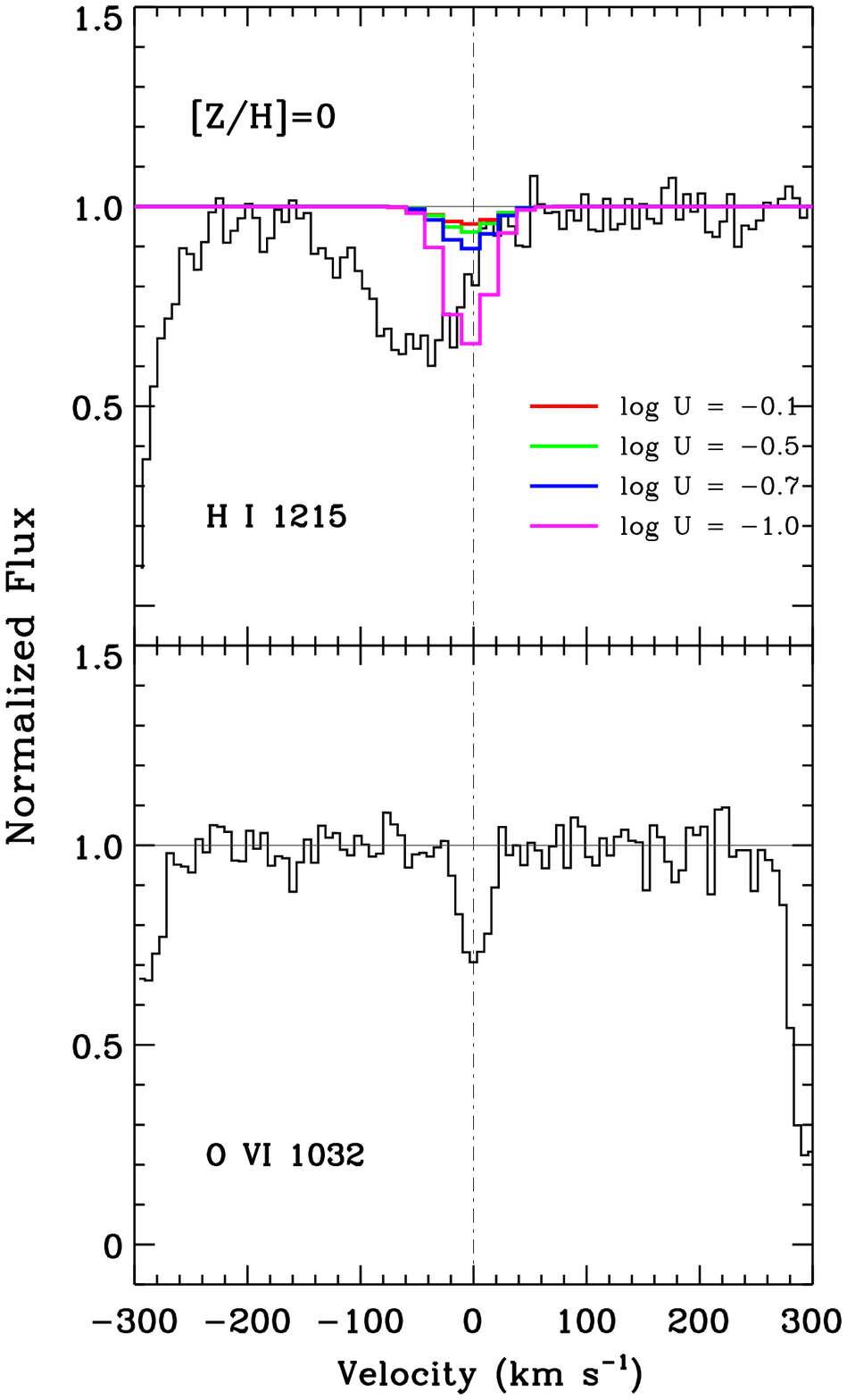}
\end{center}
\protect
\caption{\small Synthetic {\Lya} absorption profiles based on results from photoionization models for a range of log~$U$ and metallicities of 1/3 solar ({\it left panel}) and solar ({\it right panel}) are over-plotted on the data. The models were optimized to derive the measured {\OVI} column density. The predictions of the models become consistent with the {\HI} column density limit only for high values of ionization parameter such as log~$U > -0.5$. However, such high ionization parameter result in unrealistically large path length for the {\OVI} absorbing region (see Table 3).}
\label{fig:1}
\end{sidewaysfigure}

\begin{figure*}
\begin{center}
\epsscale{0.9}
\plotone{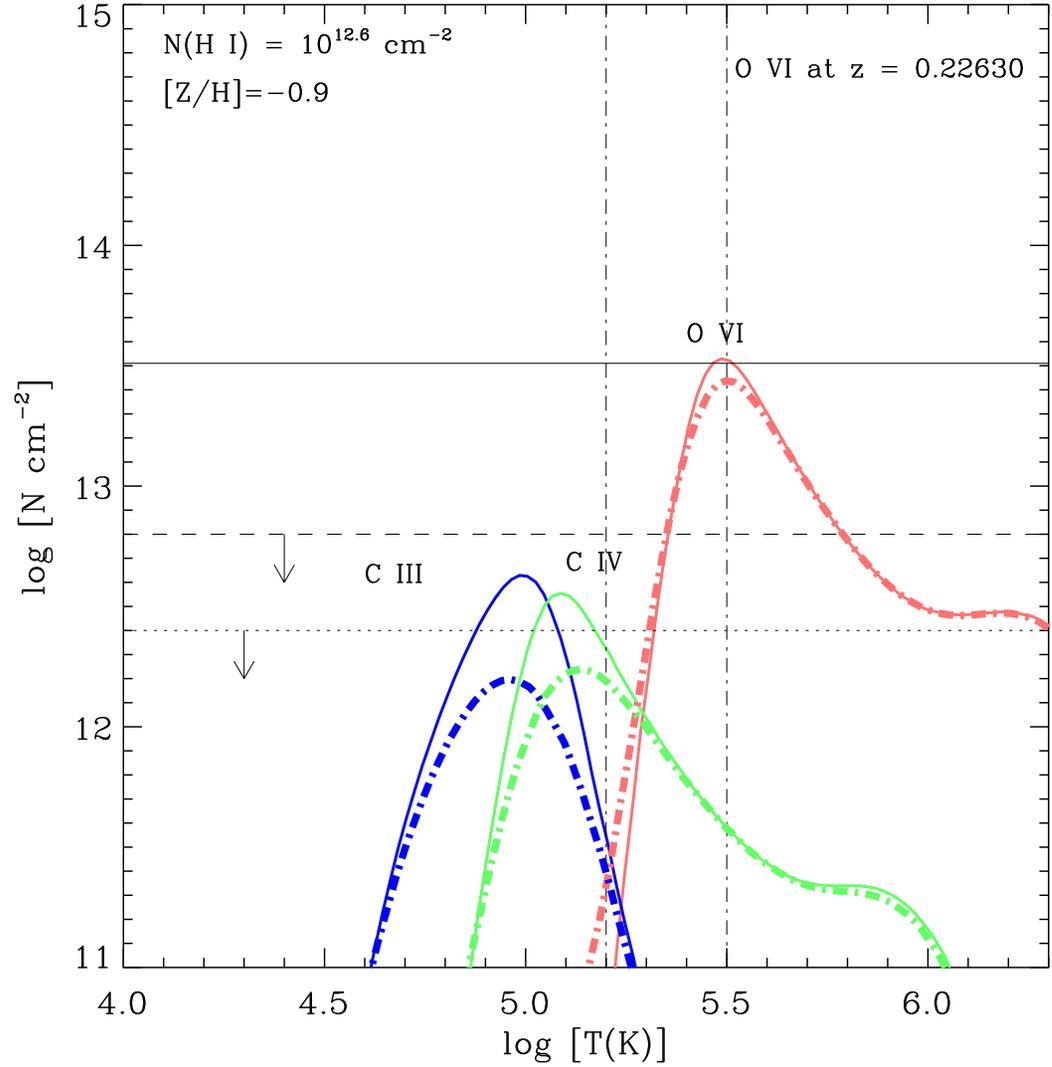}
\end{center}
\protect
\caption{\small CIE ({\it solid}) and non-CIE ({\it dash-dot}) model column density predictions for {\CIII} (blue), {\CIV} (green), and {\OVI} (red) as a function of temperature. The non-CIE model is for an isochorically cooling gas. The {\HI} column density in these models is fixed to the estimated 3$\sigma$ upper limit of $N(\HI) = 10^{12.6}$~{\cmsq} (see Table 2). The measured {\OVI} column density is represented by the horizontal {\it solid} line. The {\CIII}, and {\CIV} upper limits based on non-detection are marked using {\it dotted} and {\it dashed} lines respectively. The lower and upper temperature limits of $10^{5.3} - 10^{5.5}$~K, allowed by the data are marked by the two vertical {\it dash-dot} lines. Both CIE and non-CIE conditions appear to be consistent with the measurements. At a metallicity of -0.9~dex, the CIE and non-CIE ion fractions for {\OVI} are not significantly different from each other.}
\label{fig:1}
\end{figure*}

\begin{figure*}
\begin{center}
\epsscale{0.9}
\plotone{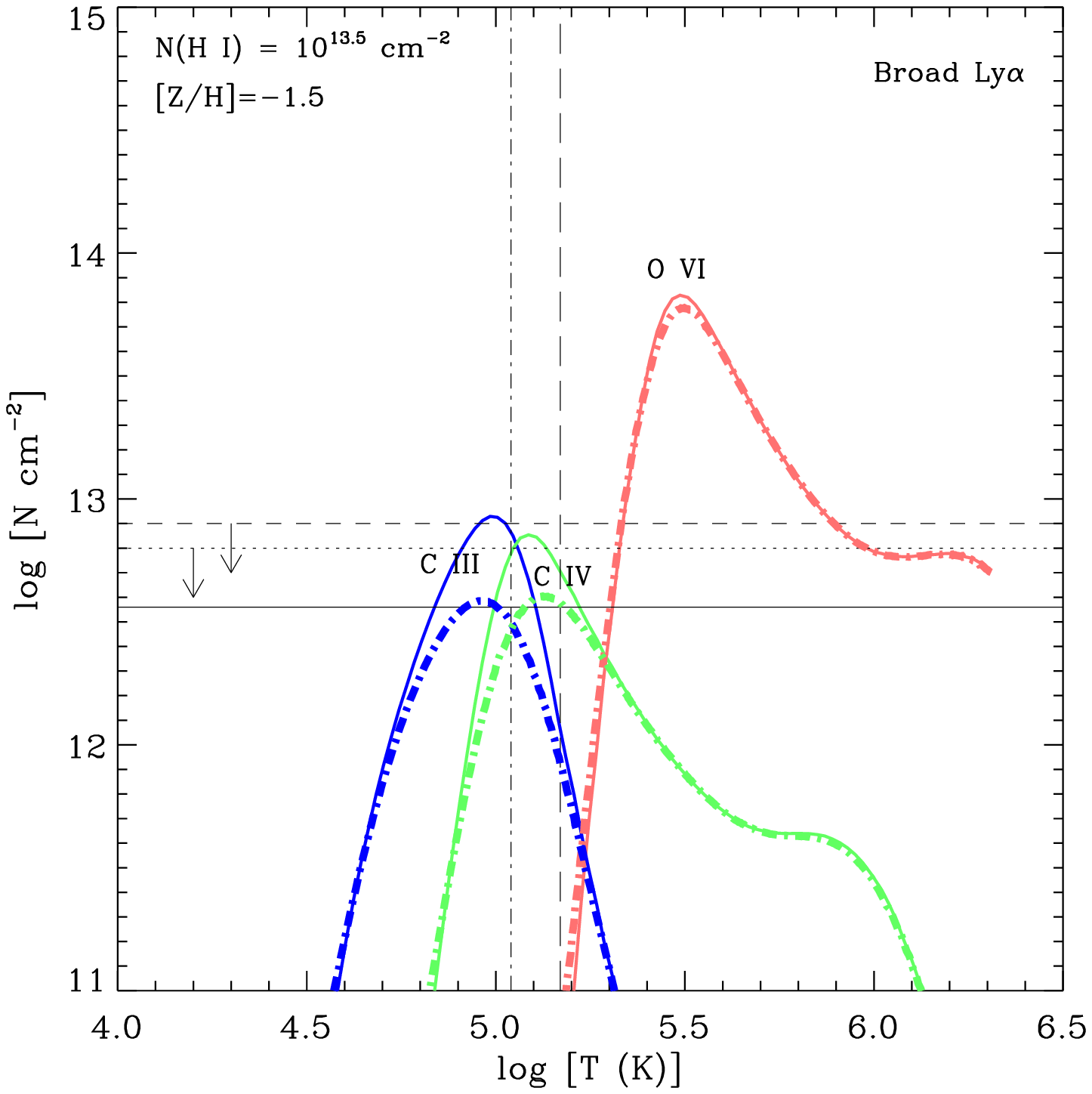}
\end{center}
\protect
\caption{\small CIE ({\it solid}) and non-CIE ({\it dash-dot}) column density predictions for {\CIII} (blue), {\CIV} (green), and {\OVI} (red) at different temperature. The non-CIE model is for an isochorically cooling gas. The {\HI} column density in these models is set to the value measured for the broad {\Lya}. The measured {\CIII} column density is marked by the horizontal {\it solid} line. The {\OVIdblt}~{\AA} and {\CIVdblt}~{\AA} lines are non-detections at the velocity of the BLA, and the corresponding upper limits are represented by the {\it dashed} and {\it dotted} lines respectively. The temperature $T = 1.1 \times 10^5$~K obtained from $b(\HI) = 51~{\pm}~2$~{\kms} and $b(\CIII) = 28~{\pm}~4$~{\kms} is marked using the vertical {\it dash-dot} line. The {\it dashed} vertical line is the upper temperature limit if the entire BLA line width is due to thermal broadening. At the temperature implied by the line widths, rapid metal line cooling introduces a $\sim 0.3$~dex difference  between the non-CIE and CIE column density predictions. From the given constraints, the non-CIE model estimates a carbon elemental abundance of [C/H] = -1.5 dex and a large baryon content of $N(\H)= 3.2 \times 10^{18}$~{\cmsq} in the BLA.}
\label{fig:1}
\end{figure*}


\clearpage
\begin{deluxetable}{lcrrrr}
\tabletypesize{\scriptsize} 
\tablewidth{0pt}
\tablecaption{\textsc{Properties of the $z = 0.22496$ {\OVI} Absorption System}}
\tablehead{
\colhead{Line} &
\colhead{$W_r$} &
\colhead{log~$N$} &
\colhead{$b$} &
\colhead{$[-v, +v]$}  & 
\colhead{Notes} \\ 
\colhead{(\AA)} &
\colhead{(m\AA)} &
\colhead{(\cmsq)} &
\colhead{(\kms)} &
\colhead{(\kms)} &
\colhead{ }
}
\startdata
{\Lya}, $\beta$, $\gamma$, $\epsilon$, $\theta$  &	$...$	&  $14.12~{\pm}~0.12$   &  $42~{\pm}~6$ &  $-96~{\pm}~9$ &  1 \\
{ }  &  $ ... $    &  $15.18~{\pm}~0.03$	 &  $26~{\pm}~2$ &  $-40~{\pm}~1$  & 1 \\
{ }  &  $ ... $    &  $15.28~{\pm}~0.03$	 &  $18~{\pm}~1$ &  $18~{\pm}~1$  & 1 \\
{ }  &  $ ... $   &  $13.05~{\pm}~0.21$	 &  $30^{+15}_{-10}$ &  $71~{\pm}~1$  & 1 \\
\\
{\Lya}, $\beta$, $\gamma$, $\delta$, $\epsilon$, $\zeta$ & $...$ & $13.87~{\pm}~1.83$   &  $32~{\pm}~14$ &  $-117~{\pm}~35$ &  2 \\
{ }  &  $ ... $    &  $15.12~{\pm}~0.04$	 &  $12~{\pm}~2$ &  $-45~{\pm}~1$  & 2\\
{ }  &  $ ... $    &  $15.10~{\pm}~0.02$	 &  $45~{\pm}~0$ &  $-23~{\pm}~1$  & 2\\
{ }  &  $ ... $    &  $15.20~{\pm}~0.02$	 &  $15~{\pm}~1$ &  $20~{\pm}~1$  & 2\\
{ }  &  $ ... $    &  $15.02~{\pm}~0.05$	 &  $13~{\pm}~1$ &  $39~{\pm}~1$  & 2\\
\\
Ly $\gamma$, $\delta$, $\epsilon$, $\zeta$ & $...$ & $15.18~{\pm}~0.01$   &  $32~{\pm}~1$ &  $-41~{\pm}~1$ &  2 \\
{ } & $ ... $ & $15.26~{\pm}~0.01$ & $23~{\pm}~1$ & $22~{\pm}~1$ & 2 \\
\\
{\Lya}  & 	  $ 847.1~{\pm}~10.7$  &  $ > 14.7$  &  $80~{\pm}~2$ &  $[-220, 130]$  & 3\\
\\
\hline
\\
{\OVIdblt}  & $ ... $ & $14.27~{\pm}~0.02$ & $45~{\pm}~2$ & $0~{\pm}~1$ & 1\\
{ }	& $ ... $  &  $13.16~{\pm}~0.11$ & $10~{\pm}~3$ & $60~{\pm}~2$ & 1\\
\\
{\OVIdblt} & $ ... $ & $13.39~{\pm}~0.29$ & $48~{\pm}~30$ & $-71~{\pm}~2$ & 2\\
{ } & $ ... $ & $13.30~{\pm}~0.16$ & $15~{\pm}~3$ & $-37~{\pm}~1$ & 2\\
{ } & $ ... $ & $14.14~{\pm}~0.03$ & $30~{\pm}~1$ & $7~{\pm}~1$ & 2\\
{ } & $ ... $ & $13.44~{\pm}~0.03$ & $15~{\pm}~1$ & $60~{\pm}~1$ & 2\\
\\
{\OVI}~$\lambda 1032$ & $... $ & $13.08~{\pm}~0.07$ & $20~{\pm}~4$ & $-86~{\pm}~2$ & 2\\
{ } & $ ... $ & $13.41~{\pm}~0.15$ & $21~{\pm}~4$ & $-37~{\pm}~1$ & 2\\
{ } & $ ... $ & $14.13~{\pm}~0.03$ & $31~{\pm}~2$ & $6~{\pm}~1$ & 2\\
{ } & $ ... $ & $13.40~{\pm}~0.03$ & $14~{\pm}~1$ & $61~{\pm}~1$ & 2\\		  
\\
{\OVI}~$\lambda 1038$ & $... $ & $13.92~{\pm}~0.03$ & $25~{\pm}~2$ & $-35~{\pm}~4$ & 2\\
{ } & $ ... $ & $13.92~{\pm}~0.09$ & $15~{\pm}~2$ & $6~{\pm}~1$ & 2\\
{ } & $ ... $ & $13.72~{\pm}~0.17$ & $18~{\pm}~7$ & $37~{\pm}~1$ & 2\\
{ } & $ ... $ & $13.22~{\pm}~0.22$ & $11~{\pm}~3$ & $65~{\pm}~2$ & 2\\		
\\
{\OVI}~$\lambda 1032$ & $183.1~{\pm}~9.0$ & $14.27~{\pm}~0.02$ & $55~{\pm}~4$ & $[-120, 100]$ & 3\\
{\OVI}~$\lambda 1038$ & $112.2~{\pm}~9.4$ & $14.32~{\pm}~0.04$ & $53~{\pm}~5$ & $[-120, 100]$ & 3\\
\\
\hline
\\
{\SiIII}~$\lambda 1207$ & $18.16~{\pm}~4.45$ & $12.05~{\pm}~0.03$ & $15~{\pm}~2$ & $-39~{\pm}~1$ & 2, 6\\
{}   & 	$50.23~{\pm}~3.35$  &  $12.51~{\pm}~0.01$ & $9~{\pm}~1$ & $3~{\pm}~1$ & 2, 6\\ 
{}   &	$32.43~{\pm}~3.00$  &  $12.33~{\pm}~0.01$ & $6~{\pm}~1$ & $31~{\pm}~1$ & 2, 6\\
\\
{\SiII}~$\lambda 1260$ & $ 4.8~{\pm}~0.8$ & $11.61~{\pm}~0.06$ & $4~{\pm}~1$ & $29~{\pm}~1$ & 2\\
\\
{\CII}~$\lambda 1335$ & $21.83~{\pm}~3.75$ & $13.08~{\pm}~0.09$ & $10~{\pm}~3$ & $5~{\pm}~0$ & 2\\
\\
{\CIII}~$\lambda 977$ & $...$ &  $13.39~{\pm}~0.03$ & $16~{\pm}~1$ & -$38~{\pm}~1$ & 2\\
{ } & $ ... $ & $13.62~{\pm}~0.04$ & $12~{\pm}~1$ & $-6~{\pm}~1$ & 2\\
{ } & $ ... $ & $13.17~{\pm}~0.05$ & $6~{\pm}~1$ & $23~{\pm}~1$ & 2\\
{ } & $ ... $ & $13.54~{\pm}~0.03$ & $64~{\pm}~3$ & $24~{\pm}~1$ & 2\\
\\
{\CIII}~$\lambda 977$ & $301.6~{\pm}~6.6$ &  $ > 14.0$ & $46~{\pm}~3$ & [-80, 100] & 3
\\
{\CIV}~$\lambda 1548$ & $126.8~{\pm}~24.9$ &  $13.58^{+0.07}_{-0.09}$ & $41~{\pm}~15$ & [-120, 100] & 3
\\
{\OIII}~$\lambda 833$ & $188.3~{\pm}~12.3$ &  $14.66^{+0.03}_{-0.03}$ & $46~{\pm}~4$ & [-70, 58] & 3
\\
{\SiIV}~$\lambda 1394$ & $8.9~{\pm}~4.6$ & $12.07~{\pm}~0.08$ & $6~{\pm}~2$ & $-35~{\pm}~1$ & 2, 5, 6\\
{} & $8.5~{\pm}~3.2$ & $11.96~{\pm}~0.09$ & $4~{\pm}~2$ & $1~{\pm}~1$ & 2, 5, 6\\
{} & $17.8~{\pm}~5.0$ & $12.28~{\pm}~0.06$ & $11~{\pm}~2$ & $33~{\pm}~1$ & 2, 5, 6\\
\\
{\SiIV}~$\lambda 1394$ & $23.5~{\pm}~24.9$ &  $< 12.9$ & $ ... $ & [-120, 100] & 4
\\
{\NV}~$\lambda 1239$ & $3.9~{\pm}~15.8$ & $< 13.3$ & $ ... $ & [-120, 100]  & 4
\\
{\NeVIII}~$\lambda 770$ & $< 56$ & $< 14.2$ & $ ... $ & [-120, 100] & 4
\enddata
\tablecomments{ The velocity errors listed are from the profile fit code. They do not include the STIS and FUSE velocity calibration errors of $\sim 3$~{\kms} and $\sim 5$~{\kms} respectively. (1) These measurements based on profile fits to the absorption lines are taken from Tripp {\etal}(2008). (2) Line parameters obtained from the profile fitting routine described in \citep{churchillthesis}. (3) Measurement based on the apparent optical depth (AOD) method of Savage \& Sembach (1991). In the case of {\Lya} and {\CIII}~$\lambda 977$~{\AA}, the lines are strong and saturated and therefore the column densities are upper limits. (4) Limits based on non-detection at the 3$\sigma$ level. (5) The {\SiIV}~$\lambda 1394$~{\AA} could be a marginal detection. Since the components are adequately resolved, we also quote the equivalent width for each component. The fit results are based on the assumption that the weak features at the respective velocities are {\SiIV}. As it is an uncertain detection, we also quote a lower limit for the column density by integrating over a velocity range. (6) The components are adequately resolved, and hence we list the equivalent width for each component.}
\label{tab:tab3}
\end{deluxetable}

\clearpage
\begin{deluxetable}{lccccr}
\tabletypesize{\scriptsize} 
\tablewidth{0pt}
\tablecaption{\textsc{Properties of the $z = 0.22638$ {\OVI} Absorption System}}
\tablehead{
\colhead{Line} &
\colhead{$W_r$} &
\colhead{log~$N$} &
\colhead{$b$} &
\colhead{$[-v, +v]$}  & 
\colhead{Notes} \\ 
\colhead{(\AA)} &
\colhead{(m\AA)} &
\colhead{(\cmsq)} &
\colhead{(\kms)} &
\colhead{(\kms)} &
\colhead{ }
}
\startdata
{\Lya}  &	$< 107.0$	&  $< 13.4$	&  $...$ &  $[-75, 75]$ &  1 \\
{\Lya}  & 	  $< 25.0$    &  $< 12.7$	 &  $...$ &  $2 \times [0, 75]$ & 2 \\ 
{\HI1025} &  $< 12.2$ & $< 13.2 $	& $...$ & $[-75, 75]$ & \\
{\OVIdblt}  &	$31.5~{\pm}~3.8$ &  $13.51~{\pm}~0.04$ &  $16~{\pm}~2$ & $0~{\pm}~1$ & 3 \\
{\OVIdblt} & $36.1~{\pm}~1.3$ & $13.51~{\pm}~0.01$ & $17~{\pm}~1$ & $0~{\pm}~1$ & 4\\
{\OVI}~$\lambda 1032$ & $27.8~{\pm}~2.5$ &  $13.40~{\pm}~0.04$ & $16~{\pm}~3$ & $[-30, 30]$ &  5\\
{\CII}~$\lambda 1335$	& $< 15.5$  &  $< 12.9$ & $...$ & $[-30, 30]$ & 6\\
{\CIII}~$\lambda 977$ & $< 15.6$ & $< 12.4$ & $...$ & $[-30, 30]$ & 6 \\
{\CIV}~$\lambda 1548$ & $< 25.7$ & $< 12.8$ & $...$ & $[-30, 30]$ & 6, 7\\
{\SiII}~$\lambda 1260$ &  $< 14.8$ & $< 12.1$ & $...$ & $[-30, 30]$ & 6\\
{\SiIII}~$\lambda 1207$ & $< 13.5$ & $< 11.9$ & $...$ & $[-30, 30]$ & 6\\
{\SiIV}~$\lambda 1403$ & $< 40.4$ & $< 12.9$ & $...$ & $[-30, 30]$ & 6, 7\\
{\NV}~$\lambda 1234$ & $< 16.9$ & $ < 12.9$ & $...$ & $[-30, 30]$ & 6\\
{\NeVIII}~$\lambda 770$ & $< 38.0$ & $< 14.0$ & $...$ & $[-30, 30]$ & 6\\
{\NeVIII}~$\lambda 780$ & $< 38.0$ & $< 14.2$ & $...$ & $[-30, 30]$ & 6 \\

\enddata
\tablecomments{The velocity errors listed are from the profile fit code. They do not include the STIS and FUSE velocity calibration errors of $\sim 3$~{\kms} and $\sim 5$~{\kms} respectively. (1) The {\HI} centered on the {\OVI} line is likely affected by blending with the stronger broad {\Lya} feature. Hence the equivalent width and column density measurements are upper limits. (2) The equivalent width was measured by integrating over the positive side of the {\OVI} line profile where there is negligible blending with the broad {\Lya} feature. The equivalent width was then doubled to get an upper limit on the measurement, and subsequently the column density was determined assuming that the line is in the linear part of the curve of growth. (3) The line parameters are fit results taken from Tripp {\etal}(2008) and obtained by a simultaneous Voigt profile fit to both members of the {\OVI} doublet. (4) Line parameters obtained by combining the Voigt profile fitting procedure AUTOVP (Dav\'e {\etal}2000) and $\chi^2$ minimization routine MINFIT (Churchill {\etal}1999). (5) Measurement based on the apparent optical depth (AOD) method of Savage \& Sembach (1991). (6) A $3\sigma$ upper limit based on non-detection. (7) Measurement based on a lower resolution $STIS$ G230M grating spectrum.}
\label{tab:tab1}
\end{deluxetable}

\begin{deluxetable}{lccccr}
\tabletypesize{\scriptsize} 
\tablewidth{0pt}
\tablecaption{\textsc{Properties of the broad {\Lya} Absorption System}}
\tablehead{
\colhead{Line} &
\colhead{$W_r$} &
\colhead{log~$N$} &
\colhead{$b$} &
\colhead{$[-v, +v]$}  & 
\colhead{Notes} \\ 
\colhead{(\AA)} &
\colhead{(m\AA)} &
\colhead{(\cmsq)} &
\colhead{(\kms)} &
\colhead{(\kms)} &
\colhead{ }
}
\startdata
{\Lya}  &	$151~{\pm}~10.8$	&  $13.52~{\pm}~0.02$	&  $56~{\pm}~4$ &  $-53~{\pm}~2$ &  1 \\
{\Lya}  & 	  $ ... $    &  $13.50~{\pm}~0.01$	 &  $51~{\pm}~2$ &  $-50~{\pm}~1$  & 2\\ 
{ }  & $ ... $    &  $12.30~{\pm}~0.01$	 &  $14~{\pm}~3$ &  $-127~{\pm}~1$ & 2 \\ 
{\Lya} &  $...$ & $13.51~{\pm}~0.02$	& $56~{\pm}~4$ & $[-160, 55]$ & 3 \\
{\OVI}~$\lambda 1032$ & $< 9.4$ &  $< 12.9$ & $ ... $ & $[-100, -35]$ &  4\\
{\CIII}~$\lambda 977$ & $17.3~{\pm}~5.4$ & $< 12.6$ & $...$ & $[-100, -35]$ & 5 \\
{\CIII}~$\lambda 977$ & $...$ & $12.56~{\pm}~0.05$ & $28~{\pm}~4$ & $-52~{\pm}~2$ & 6 \\
{\CIV}~$\lambda 1548$ & $< 31.0$ & $< 12.8$ & $...$ & $[-100, -35]$ & 4 \\
\enddata
\tablecomments{The velocity errors listed are from the profile fit code. They do not include the STIS and FUSE velocity calibration errors of $\sim 3$~{\kms} and $\sim 5$~{\kms} respectively. (1) These measurements are taken from Tripp {\etal}(2008), where a single Voigt profile is fit to the BLA absorption feature. The velocity of $v = -53~{\pm}~3$~{\kms} corresponds to the velocity offset of the BLA line centroid from $z = 0.22638$, the redshift of the {\OVI} absorber. (2) Line parameters obtained by combining the Voigt profile fitting procedure AUTOVP (Dav\'e {\etal}2000) and $\chi^2$ minimization routine MINFIT (Churchill {\etal}1999). The {\Lya} profile is resolved into two components, where one of the components is a BLA (see Figure 1). (3) Measurement based on the apparent optical depth (AOD) method of \citet{savage91}. (4) Upper limit derived by integrating the wavelength region corresponding to the redshifted {\OVI}~$\lambda 1032$~{\AA} line over the velocity interval given in the table. The column density limit was derived assuming that the measurement is in the linear part of the curve of growth. (5) Line parameters obtained by Voigt profile fitting. The velocity corresponding to the centroid of the profile fit, is within 1$\sigma$ of the BLA line centroid.}
\label{tab:tab3}
\end{deluxetable}

\clearpage
\begin{landscape}
\begin{deluxetable}{lccccccccccr}
\tabletypesize{\scriptsize} 
\tablewidth{0pt}
\tablecaption{\textsc{Summary of the photoionization modeling results for the $z = 0.22638$ {\OVI} Absorption}}
\tablehead{
\colhead{[Z/H]} &
\colhead{$U$} &
\colhead{$N(\H)$} &
\colhead{$N(\HI)$} &
\colhead{$N(\OVI)$}  & 
\colhead{$N(\CIII)$} &
\colhead{$N(\CIV)$} &
\colhead{$n_{\H}$} &
\colhead{T} &
\colhead{$p/k$} &
\colhead{L} &
\colhead{Comments} \\
\colhead{(dex)} &
\colhead{(dex)} &
\colhead{(dex)} &
\colhead{(dex)} &
\colhead{(dex)}  & 
\colhead{(dex)}  & 
\colhead{(dex)} &
 \colhead{(\cc)} &
\colhead{(K)} &
\colhead{(K~{\cc})} &
\colhead{(kpc)} &
\colhead{ }
}
\startdata
$-0.5$ &  $-2.0$ & $20.40$ & $16.97$ & $13.51$ & $16.20$ & $15.18$ & $1.88 \times 10^{-4}$ & $1.40 \times 10^4$ & $2.63$ & $434$ & {\HI}, {\CIII}, {\CIV} over produced, large path length \\
$-0.5$ & $-1.0$ &	$18.25$ & $13.65$ & $13.51$ & $13.21$ & $13.19$ & $1.88 \times 10^{-5}$ & $2.80 \times 10^4$ & $0.53$ & $31$ & {\HI}, {\CIII}, {\CIV} over produced \\
$-0.5$ & $-0.7$ &	$18.10$ & $13.12$ & $13.51$ & $12.43$ & $12.65$ & $9.41 \times 10^{-6}$ & $3.30 \times 10^4$ & $0.31$ & $43$ & {\HI} over produced \\
$-0.5$ & $-0.5$ & $18.10$ & $12.90$ &  $13.51$ & $11.94$ & $12.32$ & $5.93 \times 10^{-6}$ & $3.83 \times 10^4$ & $0.23$ & $69$ & {\HI} over produced \\
$-0.5$ & $-0.3$ & $18.25$ & $12.79$ & $13.51$ & $11.47$ & $12.02$ & $3.74 \times 10^{-6}$ & $4.49 \times 10^4$ & $0.17$ & $154$ & large path length \\
$-0.5$ & $-0.1$ & $18.50$ & $12.79$ & $13.51$ & $11.01$ & $11.74$ & $2.36 \times 10^{-6}$ & $1.40 \times 10^4$ & $0.03$ & $433$ & large path length \\
\\
\hline
\\
$-0.3$ & $-2.0$ & $20.12$ & $16.73$ & $13.51$ & $16.12$ & $15.12$ & $1.88 \times 10^{-4}$ & $1.25 \times 10^4$ & $2.35$ & $228$ & {\HI}, {\CIII}, {\CIV} over produced, large path length \\
$-0.3$ & $-1.0$ & $18.03$ & $13.46$ & $13.51$ & $13.18$ & $13.22$ & $1.88 \times 10^{-5}$ & $2.52 \times 10^4$ & $0.47$ & $19$ & {\HI}, {\CIII}, {\CIV} over produced \\
$-0.3$ & $-0.7$ & $17.84$ & $12.91$ & $13.51$ & $12.41$ & $12.66$ & $9.41 \times 10^{-6}$ & $3.08 \times 10^4$ & $0.29$ & $24$ & {\HI} over produced \\
$-0.3$ & $-0.5$ & $17.87$ & $12.69$ & $13.51$ & $11.93$ & $12.33$ & $5.93 \times 10^{-6}$ & $3.58 \times 10^4$ & $0.21$ & $41$ & {\HI} over produced \\
$-0.3$ & $-0.3$ & $18.02$ & $12.58$ & $13.51$ & $11.47$ & $12.02$ & $3.74 \times 10^{-6}$ & $4.30 \times 10^4$ & $0.16$ & $91$ & large path length \\
$-0.3$ & $-0.1$ & $18.30$ & $12.58$ & $13.51$ & $11.00$ & $11.73$ & $2.36 \times 10^{-6}$ & $5.33 \times 10^4$ & $0.13$ & $274$ & large path length \\
\\
\hline
\\
$0.0$ & $-2.0$ & $19.73$ & $16.41$ & $13.51$ & $16.03$ & $15.01$ & $1.88 \times 10^{-4}$ & $1.40 \times 10^4$ & $2.63$ & $93$ & {\HI}, {\CIII}, {\CIV} over produced, large path length \\
$0.0$ & $-1.0$ & $17.67$ & $13.16$ & $13.51$ & $13.13$ & $13.24$ & $1.88 \times 10^{-5}$ & $2.17 \times 10^4$ & $0.41$ & $8$ & {\HI}, {\CIII}, {\CIV} over produced \\
$0.0$ & $-0.7$ & $17.50$ & $12.61$ & $13.51$ & $12.38$ & $12.68$ & $9.41 \times 10^{-6}$ & $2.76 \times 10^4$ & $0.26$ & $11$ & acceptable, but marginal \\
$0.0$ & $-0.5$ & $17.55$ & $12.40$ & $13.51$ & $11.93$ & $12.36$ & $5.93 \times 10^{-6}$ & $3.29 \times 10^4$ & $0.20$ & $19$ & acceptable \\
$0.0$ & $-0.3$ & $17.70$ & $12.28$ & $13.51$ & $11.47$ & $12.03$ & $3.74 \times 10^{-6}$ & $4.09 \times 10^4$ & $0.15$ & $43$ & acceptable \\
$0.0$ & $-0.1$ & $18.02$ & $12.30$ & $13.51$ & $11.02$ & $11.75$ & $2.36 \times 10^{-6}$ & $5.40 \times 10^4$ & $0.13$ & $144$ & large path length \\
\enddata
\tablecomments{The predicted model column densities and other physical parameters at different metallicities and ionization parameters in order to recover the observed value of {\OVI} column density under photoionization equilibrium conditions. The number density of ionizing photons with $h\nu \geq 13.6$~eV is estimated from the Haardt \& Madau (2001) ionizing background radiation field to be $n_\gamma = 1.88 \times 10^{-6}$~{\cc}. The elemental abundance for C and O are assumed to be solar, with  [C/H] = -3.61 and [O/H] = -3.31 as described in Cloudy (ver. 08) and Allende Prieto {\etal}(2001, 2002). }
\label{tab:tab2}
\end{deluxetable}
\clearpage
\end{landscape}

\clearpage
\begin{landscape}
\begin{deluxetable}{lccccccccccr}
\tabletypesize{\scriptsize} 
\tablewidth{0pt}
\tablecaption{\textsc{Summary of the photoionization modeling results for the broad {\Lya} Absorption}}
\tablehead{
\colhead{[Z/H]} &
\colhead{$U$} &
\colhead{$N(\H)$} &
\colhead{$N(\HI)$} &
\colhead{$N(\OVI)$}  & 
\colhead{$N(\CIII)$} &
\colhead{$N(\CIV)$} &
\colhead{$n_{\H}$} &
\colhead{T} &
\colhead{$p/k$} &
\colhead{L} &
\colhead{Comments} \\
\colhead{(dex)} &
\colhead{(dex)} &
\colhead{(dex)} &
\colhead{(dex)} &
\colhead{(dex)}  & 
\colhead{(dex)}  & 
\colhead{(dex)} &
 \colhead{(\cc)} &
\colhead{(K)} &
\colhead{(K~{\cc})} &
\colhead{(kpc)} &
\colhead{ }
}
\startdata
$-0.5$ & $-5.0$ & $14.04$ & $13.50$ & $ 0$ & $ 8.44$ & $3.95$ & $1.88 \times 10^{-1}$ & $7.60 \times 10^3~^\dagger $ & $1430.00$ & $1.89 \times 10^{-7}$ & {\CIII} under produced \\
$-0.5$ & $-4.0$ & $14.83$ & $13.50$ & $ 0.50$ & $10.07$ & $6.67$ & $1.88 \times 10^{-2}$ & $9.16 \times 10^3~^\dagger $ & $172.00$ & $1.17 \times 10^{-5}$ & {\CIII} under produced \\
$-0.5$ & $-3.0$ & $15.85$ & $13.50$ & $ 6.58$ & $11.61$ & $9.64$ & $1.88 \times 10^{-3}$ & $1.09 \times 10^4~^\dagger $ & $20.50$ & $0.001$ & {\CIII} under produced \\
$-0.5$ & $-2.0$ & $16.97$ & $13.50$ & $10.62$ & $12.74$ & $11.95$ & $1.88 \times 10^{-4}$ & $1.67 \times 10^4~^\dagger $ & $3.13$ & $0.161$ & {\CIII} mildly over produced in the wings \\
$-0.5$ & $-1.5$ & $17.55$ & $13.50$ & $12.15$ & $13.09$ & $12.70$ & $5.93 \times 10^{-5}$ & $2.17 \times 10^4~^\dagger $ & $1.29$ & $1.938$ & {\CIII} over produced \\
$-0.5$ & $-1.0$ & $18.13$ & $13.50$ & $13.39$ & $13.08$ & $13.07$ & $1.88 \times 10^{-5}$ & $2.80 \times 10^4~^\dagger $ & $0.53$ & $23.295$ & {\CIII}, {\CIV}, {\OVI} over produced\\
\\
\hline
\\
$-1.0$ & $-5.0$ & $14.09$ & $13.50$ & $ 0$ & $ 8.02$ & $3.52$ & $1.88 \times 10^{-1}$ & $9.44 \times 10^3~^\dagger $ & $1771.55$ & $2.12 \times 10^{-7}$ & {\CIII} under produced \\
$-1.0$ & $-4.0$ & $14.89$ & $13.50$ & $ 0.11$ & $9.66$ & $6.29$ & $1.88 \times 10^{-2}$ & $1.13 \times 10^4~^\dagger $ & $212.06$ & $1.34 \times 10^{-5}$ & {\CIII} under produced \\
$-1.0$ & $-3.0$ & $15.92$ & $13.50$ & $ 6.27$ & $11.19$ & $9.28$ & $1.88 \times 10^{-3}$ & $1.41 \times 10^4~^\dagger $ & $26.46$ & $0.001$ & {\CIII} under produced \\
$-1.0$ & $-2.0$ & $17.06$ & $13.50$ & $10.24$ & $12.32$ & $11.44$ & $1.88 \times 10^{-4}$ & $2.21 \times 10^4~^\dagger $ & $4.15$ & $0.198$ & {\CIII} under produced \\
$-1.0$ & $-1.5$ & $17.63$ & $13.50$ & $11.67$ & $12.66$ & $12.15$ & $5.93 \times 10^{-5}$ & $2.86 \times 10^4~^\dagger $ & $1.69$ & $2.330$ & {\CIII} mildly over produced \\
$-1.0$ & $-1.0$ & $18.20$ & $13.50$ & $12.88$ & $12.63$ & $12.53$ & $1.88 \times 10^{-5}$ & $3.59 \times 10^4~^\dagger $ & $0.67$ & $27.370$ & {\CIII} over produced\\
\enddata
\tablecomments{$^\dagger$ The photoionization equilibrium temperature is a factor of $\sim 4 - 15$ lower than the minimum temperature estimated ($T = 1.1 \times 10^5$~K) from the Doppler line widths of {\Lya} and {\CIII}~$\lambda 977$~{\AA} lines. Even for the model that is approximately consistent with the constrains set by the {\CIII} and {\OVI} column densities, the derived model temperature is a factor of 7 smaller than the minimum temperature derived from the line widths. A $T \sim 10^5$~K cannot be recovered from a gas phase in photoionization equilibrium.}
\label{tab:tab4}
\end{deluxetable}
\clearpage
\end{landscape}

\end{document}